\documentclass[a4paper,11pt]{article}
\pdfoutput=1 

\usepackage{jheppub} 
\usepackage{bbm}

\newcommand\cm{\mathcal{M}}

\newcommand\beal{\begin{align}}
\newcommand\nn{\nonumber}
\newcommand\bbone{\ensuremath{\mathbbm{1}}}

\newcommand{\eq}[1]{\begin{equation}#1\end{equation}}
\newcommand{\spl}[1]{\begin{split}#1\end{split}}
\newcommand{\al}[1]{\begin{align}#1\end{align}}

\newcommand{\mcal}{\mathcal{M}}

\newcommand{\ncal}{\mathcal{N}}

\newcommand{\G}{\Gamma}
\newcommand{\g}{\gamma}
\newcommand{\e}{\epsilon}
\newcommand{\we}{\widetilde{\eta}}

\newcommand{\p}{\partial}
\renewcommand{\O}{\Omega}

\renewcommand{\a}{\alpha}
\renewcommand{\b}{\beta}
\renewcommand{\o}{\omega}
\renewcommand{\t}{\theta}

\def\d{\text{d}}

\def\slashchar#1{\setbox0=\hbox{$#1$}           
\dimen0=\wd0                                 
\setbox1=\hbox{/} \dimen1=\wd1               
\ifdim\dimen0>\dimen1                        
\rlap{\hbox to \dimen0{\hfil/\hfil}}      
#1                                        
\else                                        
\rlap{\hbox to \dimen1{\hfil$#1$\hfil}}   
/                                         
\fi}

\title{IIA supergravity and M-theory on manifolds with $SU(4)$ structure}

\author{Dani\"{e}l Prins and}
\author{Dimitrios Tsimpis}

\affiliation{Universit\'{e} de Lyon\\
UMR 5822, CNRS/IN2P3, Institut de Physique Nucl\'{e}aire de Lyon\\
4 rue Enrico Fermi, 69622 Villeurbanne Cedex,  France\\}
\emailAdd{dlaprins@ipnl.in2p3.fr}
\emailAdd{tsimpis@ipnl.in2p3.fr}

\abstract{We give the general form of supersymmetric backgrounds with two real supercharges of M-theory and type IIA supergravity (with non-zero Romans mass in general)  of the form $\mathbb{R}^{1,d}\times \mcal_8$, $d=1,2$, on eight-dimensional manifolds with $SU(4)$ structure. We point out a  subtlety in the integrability theorems for low-dimensional supersymmetric compactifications. As a special case we examine Calabi-Yau flux vacua and we show that unbroken supersymmetry does {\em not} in general require the four-form flux to be (2,2) or primitive. Our results could be used to construct novel higher-dimensional analogues of the Klebanov-Strassler geometry. In the case of M-theory large-volume Calabi-Yau flux vacua our results are in agreement with partial supersymmetry breaking in three-dimensional $\ncal=2$ supergravity. Alternatively, the conditions for supersymmetry can be expressed in terms of a real `superpotential' in accordance with three-dimensional $\ncal=1$ supergravity. We present explicit examples of M-theory flux vacua on  K3$\times$K3, which however do not appear to possess F-theory duals with four-dimensional Poincar\'e invariance.}

\begin{document}
\maketitle
\flushbottom

\setcounter{footnote}{0}
\renewcommand{\thefootnote}{\arabic{footnote}}
\setcounter{section}{0}

\section{Introduction}\label{introduction}\label{sec1}

Supersymmetric compactifications  of M-/string theory on eight-dimensional manifolds $\mcal_8$ with four real supercharges \cite{beck,gvw,drs,gukov} lie at the intersection of several interesting ideas:
M-theory (and the IIA string theory related to it by circle reduction) on elliptically fibered Calabi-Yau fourfolds has a dual F-theoretic formulation whereby three-dimensional $\ncal=2$ vacua of M theory correspond to four-dimensional $\ncal=1$ vacua of F-theory \cite{vafa}.
Since such backgrounds are typically warped, they can be related to
 Randall-Sundrum \cite{rsa,rsb} and Klebanov-Strassler \cite{ksa,ksb} type scenarios; they can lead to hierarchies of scales \cite{hierarchy} and the stabilization of (some of) the geometric moduli, see e.g. \cite{denef} for a review.

Compactifications of M-theory or IIA string theory on eight-manifolds with two real supercharges are relatively less well-explored. Three-dimensional $\ncal=1$ compactifications of M-theory on $Spin(7)$-holonomy manifolds are believed to be dual to non-supersymmetric four-dimensional F-theory compactifications \cite{vafa} and could clearly be of interest for phenomenology. A concrete proposal of this scenario  has recently appeared in \cite{grimm}.

At lowest order in the large-volume expansion, the eight-dimensional vacuum manifold $\mcal_8$ is constrained to be Ricci-flat \cite{beckersusybr}. For $\mcal_8$ a Calabi-Yau fourfold, the low-energy effective theory obtained  from the Kaluza-Klein reduction of M-theory on $\mcal_8$ is given by a three-dimensional $\ncal=2$ supergravity \cite{gvw,hl}. Similarly for $\mcal_8$ a $Spin(7)$-holonomy manifold, the low-energy effective theory is a three-dimensional $\ncal=1$ supergravity \cite{betal}. Analogous results hold for the Kaluza-Klein reduction of IIA string theory on $\mcal_8$ \cite{gateswitten,hg}.
On the other hand, M-theory or IIA string theory on $\mcal_8$'s which are not Ricci-flat can be thought of as arising from the backreaction of  branes wrapping internal cycles of the compactification manifold \cite{mb}.

Various aspects of $\ncal=1$ compactifications of M-theory on $Spin(7)$-holonomy manifolds have been considered in \cite{beckSpin7, gukovsparks, gukovSpin7}.
More generally, the constraints of $\ncal=1$ supersymmetry on the
local geometry of $\mcal_8$ have been formulated in terms of $G_2$ structures in \cite{mb} and $Spin(7)$ structures in \cite{tsimpism}. However supersymmetry does not in general imply the reduction of the structure group of $\mcal_8$;  rather it implies the reduction of the structure group to $Spin(7)$ of an auxiliary nine-manifold $\mcal_8\times S^1$ \cite{tsimpism}. This point has been further developed in \cite{lazb}.

In the present paper  we focus on backgrounds of M-theory and  (massive) IIA supergravity with two real supercharges  of the form $\mathbb{R}^{1,d}\times \mcal_8$, $d=1,2$, where $\mcal_8$ is a manifold with $SU(4)$ structure. Furthermore we impose a strict $SU(4)$ ansatz for the
Killing spinors. The reformulation of supersymmetry in terms of a strict $SU(4)$ structure and the associated torsion classes had never been explicitly given before for this class of backgrounds\footnote{The conditions for $\ncal=1$ M-theory backgrounds we consider here are special cases of those in \cite{tsimpism} since $SU(4)$ is a subgroup of $Spin(7)$.}. In the absence of backreacting  branes wrapping internal cycles, the vacuum manifold $\mcal_8$ is constrained to be a Calabi-Yau fourfold. We analyze explicitly this special case and we discover new possibilities for the fluxes  which go beyond the results of \cite{beck,gukov}. In particular for M-theory compactifications on Calabi-Yau fourfolds, it is sometimes stated in the literature that unbroken supersymmetry requires the four-form flux to be (2,2) and primitive. This is not true in general: as we will show, a certain combination of non-primitive (2,2) flux and (4,0) flux (and its complex conjugate) can preserve two!
  real supercharges ($\ncal=1$ in three dimensions). These results are in accordance with the findings of \cite{haacksamt}: these authors have shown the existence of spontaneously broken $\ncal=1$ vacua of three-dimensional $\ncal=2$ gauged supergravity which would uplift precisely to the supersymmetric Calabi-Yau flux vacua with two real supercharges considered here. Furthermore, in the case of manifolds with holonomy a proper subgroup of $SU(4)$, a non-primitive (3,1) flux (and its complex conjugate) is also allowed.

The outline of the remainder of this paper is as follows: in section \ref{sec2}
we start by considering type IIA vacua on manifolds $\mcal_8$ with $SU(4)$ structure. We specialize to the case of strict $SU(4)$ structure and give the general form of the solution to the Killing spinor equations with two-dimensional $\ncal=(1,1)$ supersymmetry (two real supercharges) in section \ref{medium}. In section \ref{sec:int} we point out a  subtlety concerning the integrability theorems for low-dimensional supersymmetric compactifications.

In section \ref{sec:iiath} we turn to the special case where $\mcal_8$ is a Calabi-Yau fourfold. Using  the integrability theorem of section \ref{sec:int} we derive the necessary and sufficient conditions for an $\ncal=(1,1)$ background to satisfy all the equations of motion of IIA supergravity. These results (and their uplift to M-theory) could serve as starting point for
constructing novel higher-dimensional backgrounds of Klebanov-Strassler type.

In section \ref{sec:mth} we consider M-theory vacua on manifolds with $SU(4)$ structure  with three-dimensional $\ncal=1$ supersymmetry (two real supercharges). The solution to the Killing spinor equations is obtained both by uplifting the IIA results of section \ref{medium} (after taking the limit of zero Romans mass) and by direct computation in eleven dimensions.

We specialize to the case of supersymmetric Calabi-Yau flux vacua in section \ref{sec:cy} and compare with the known results concerning  large-volume vacua \cite{beckersusybr}. We reformulate the supersymmetry conditions in this case in terms of two superpotentials, recovering the three-dimensional $\ncal=2$ supergravity vacua with partially broken $\ncal=1$ supersymmetry of \cite{haacksamt}. Furthermore we show that these vacua can also be described as $\ncal=1$ solutions of three-dimensional $\ncal=1$ supergravity, in terms of one real `superpotential'.

In section \ref{sec:k3} we specialize to $\ncal=1$,  M-theory  vacua for $\mcal_8$ a K3$\times$K3 manifold. As we show in section \ref{sec:enhancement} in this case supersymmetry can be enhanced to $\ncal=2$ or   $\ncal=4$. We give three explicit examples in section \ref{sec:examples} with $\ncal=1,2,4$ supersymmetry. We conclude in section \ref{secconcl} with some open questions. Several useful technical results are included in the appendices.

\section{Supersymmetric IIA backgrounds with $SU(4)$ structure}\label{sec2}

We will consider ten-dimensional type IIA backgrounds which
are topologically  direct products
of the form $\cm_{2}\times\cm_{8}$. The  `internal' manifold $\cm_{8}$  is assumed to be Riemannian and spin; it need not be compact. We will take the `external' manifold $\cm_{2}$ to be either $\mathbb{R}^{1,1}$ or AdS$_2$; however, as we will see in the following, our supersymmetric ansatz excludes the case of AdS$_2$, so that $\cm_{2}=\mathbb{R}^{1,1}$.
The ten-dimensional metric reads:
\eq{\label{10dmetric}
\d s^2 = e^{2A}\d s^2(\cm_2)+\d s^2(\cm_{8})
~,}
where the warp factor $A$ is taken to depend only on
the coordinates of the internal manifold $\cm_{8}$.
 The most general ansatz for the RR fluxes respecting the two-dimensional Poincar\'e symmetry of our setup are of the form:\footnote{We follow the `democratic' supergravity conventions of \cite{march}, see appendix A therein, except for the ten-dimensional Hodge star operator $*_{10}$ which we define as
\eq{\spl{\label{star10}
\star_{10}\omega_{p}\, =\, \frac{1}{p!(10-p)!}\sqrt{-g}\,\epsilon_{M_1\ldots M_{10}}\omega^{M_{11-p}\ldots M_{10}} \d x^{M_1}\wedge \ldots\wedge \d x^{M_{10-p}}\nonumber~,
}}
with $\epsilon_{01\ldots 9}=1$. }
\eq{\label{fluxan}
F^{\mathrm{tot}}=\mathrm{vol}_2\wedge F^{\mathrm{el}}+F
~,}
where $\mathrm{vol}_2$ is the unwarped volume element of $\cm_2$,
 and we are using polyform notation.
We denote by $F$ the `magnetic' RR fluxes with indices
along the internal space $\cm_{8}$.
The ten-dimensional Hodge duality relates $F$ to the `electric' RR fluxes
via:
\eq{\label{em}
F^{\mathrm{el}}=e^{2A}\star_{8}\sigma(F)
~,}
where the Hodge star above is with respect to the internal metric, and
the involution $\sigma$ acts by inverting the order of the form indices.

We will consider backgrounds which are such that $\mcal_8$ admits a reduction of its structure group to $SU(4)$. As explained in appendix \ref{sec3}, this is equivalent to the existence on $\mcal_8$ of a nowhere-vanishing pure spinor $\eta$, which we can take to be of unit norm and of positive chirality without loss of generality. The reduction of the structure group to $SU(4)$ is also  equivalent to the existence on $\mcal_8$ of an almost complex structure (which implies the reduction the structure group to $U(4)$) for which in addition $c_1=0$, where $c_n$ denotes the $n$-th Chern class of the $+i$-eigenbundle of the almost complex structure.

On the other hand, the topological obstruction to the
existence of a nowhere-vanishing Majorana-Weyl spinor on $\mcal_8$ is known
\cite{gray, isham}: it is equivalent
to the condition that the Euler characteristic of $\mcal_8$ is
given by
\beal
\chi(\mcal_8)=\frac{1}{2}\int_{\mcal_8}(p_2-\frac{1}{4}p_1^2)~,
\label{top}
\end{align}
where $p_{1,2}$ are the first and second Pontrjagin forms.
These are also given by $p_1=c_1^2-2c_2$, $p_2=c_2^2+2c_4-2c_1c_3$, see e.g. \cite{egh}, so that for $c_1=0$  (\ref{top}) becomes an identity. Hence the existence on $\mcal_8$ of an almost complex structure with vanishing first Chern class is consistent with (\ref{top}), as it should.

We will further specialize to a supersymmetric ansatz with strict $SU(4)$ structure. This means that the internal part of the Killing spinor of IIA supergravity should be a linear combination with scalar coefficients of $\eta$ and its complex conjugate. Finally we demand that the ten-dimensional backgrounds allow for kappa-symmetric branes. This condition can be expressed as the equality of the norms of the positive- and negative-chirality projections of the IIA Killing spinor.

Let us examine the most general spinor decompositions satisfying the  conditions of the previous paragraph. The IIA Killing spinor decomposes into two  Majorana-Weyl spinors  of $Spin(1,9)$ $\epsilon_{1}$, $\epsilon_{2}$ of positive, negative chirality respectively. The background possesses
$\mathcal{N}=(p,q)$ supersymmetry if the spinor ansatz is parameterized by
$p$, $q$ linearly independent Majorana-Weyl spinors $\zeta_{i+}$, $\zeta_{j-}$, $i=1,\dots, p$, $j=1,\dots, q$, of $\mcal_2$ of positive, negative chirality repsectively. Since
a Majorana-Weyl spinor of $Spin(1,1)$ has one real degree of freedom, $\mathcal{N} = (p,q)$ corresponds to p+q real supercharges. The maximal supersymmetry permitted by our ansatz corresponds to four real supercharges;\footnote{More than four real supercharges may be possible in the case where the structure group reduces to a strict subgroup of $SU(4)$. This will be further discussed in section \ref{sec:k3}.} furthermore the condition for kappa-symmetric branes excludes the cases $\mathcal{N} = (p,0)$ and $\mathcal{N} = (0,q)$. The maximally-supersymmetric case
$\mathcal{N} = (2,2)$ has been analyzed in \cite{gukov} and is reviewed in
appendix \ref{maximal} in order to make contact with our notation/conventions.
$\mathcal{N} = (1,1)$ (two real supercharges) is the case of interest for the purposes of our paper; it is analyzed in section \ref{medium}. $\mathcal{N}=(1,0)$ is the case of minimal supersymmetry; although it is excluded by the condition for kappa-symmetric branes,  for completeness
 we have analyzed it in appendix \ref{minimal}.

\subsection{$\ncal=(1,1)$}\label{medium}

This is the case of interest for the purposes of our paper.
As follows from the discussion in section \ref{sec2}, we can always
choose the phase of $\eta$ so that the Killing spinor ansatz  reads:
\eq{\label{spindecompIIA}
\epsilon_1=\frac{\alpha}{\sqrt{2}}\zeta_+\otimes(\eta+\eta^c)~,~~~
\epsilon_2=\frac{\alpha}{\sqrt{2}}\zeta_-\otimes(e^{i\t}\eta+e^{-i\t}\eta^c)~,
}
where $\alpha$, $\t$ are real functions on $\mcal_8$. The
$\zeta_+$, $\zeta_-$ are Majorana-Weyl spinors of positive, negative chirality respectively, so that $\psi\equiv\zeta_+ + \zeta_-$  is a Majorana Killing spinor of $\mathcal{M}_2$:
\eq{\label{29}
\nabla_\mu\psi=W\gamma_\mu\psi~.
}
The real constant $W$ is related to the scalar curvature $R$ of $\mathcal{M}_2$ through
\eq{R=-8W^2~,}
hence $\mathcal{M}_2$ must be of constant negative scalar scurvature. The Minkowski space $\mcal_2=\mathbb{R}^{1,1}$ is recovered in the $W\rightarrow 0$ limit; in this case both $\zeta_{\pm}$ are separately Majorana-Weyl Killing spinors of  $\mathbb{R}^{1,1}$.

In our conventions the type IIA Killing spinor equations are given by:
\eq{\spl{\label{kse}
\delta \lambda^1 &= \Big(\underline{\p}\phi + \frac{1}{2} \underline{H} \Big) \epsilon_1- \frac{1}{16} e^{\phi} \G^M \underline{F}^{\mathrm{tot}}\G_M \epsilon_2 = 0 \\
\delta \lambda^2 &= \Big( \underline{\p }\phi - \frac{1}{2} \underline{H} \Big) \epsilon_2 - \frac{1}{16} e^{\phi} \G^M \sigma(\underline{F}^{\mathrm{tot}})\G_M \epsilon_1 = 0 \\
\delta \psi^1_M &= \Big( \nabla_M + \frac{1}{4} \underline{H}_M \Big) \epsilon_1 -\frac{1}{16} e^{\phi} \underline{F}^{\mathrm{tot}}\G_M
\epsilon_2 = 0 \\
\delta \psi^2_M &= \Big( \nabla_M - \frac{1}{4} \underline{H}_M \Big) \epsilon_2 -  \frac{1}{16} e^{\phi} \sigma(\underline{F}^{\mathrm{tot}})\G_M  \epsilon_1 = 0 \;,
}}
where for any $(p+q)$-form $S$ we define:
\eq{
\underline{S}_{M_1\dots M_q}\equiv\frac{1}{p!}\Gamma^{N_1\dots N_p}
{S}_{N_1\dots N_pM_1\dots M_q}
~.}
We then decompose all fluxes into $su(4)$ modules using \eqref{fluxan} and the formul\ae{} of section \ref{sec4}. We also decompose all gamma matrices as in appendix \ref{app1} and we use the ten-dimensional Killing spinor ansatz  (\ref{spindecompIIA}). We thus obtain the following set of equations:
\eq{\spl{\label{susyiian}
W&=0\\
 2\d A+h_{1}&=0\\
\left(2\underline{\d\phi}+\underline{h_1}+\underline{h_3}\right)(\eta+\eta^c)&=0\\
\left(2\underline{\d\phi}+\underline{h_1}-\underline{h_3}\right)
\left(e^{i\t}\eta+e^{-i\t}\eta^c \right)&=0\\
\left(\nabla_m+\partial_m\log\alpha+\frac{1}{4}h_{1|m}
+\frac{1}{4}\underline{h_{3|m}}\right)(\eta+\eta^c)&=0\\
\left(\nabla_m+\partial_m\log\alpha+\frac{1}{4}h_{1|m}
-\frac{1}{4}\underline{h_{3|m}}\right)\left(e^{i\t}\eta+e^{-i\t}\eta^c \right)&=0
~,}}
and
\eq{\spl{\label{susyiiar}
{\underline{\sigma(F)}}\left(\eta+\eta^c \right) &=0\\
\underline{\sigma(F)}\gamma_m\left(\eta+\eta^c \right)&=0\\
{\underline{F}}\left(e^{i\t}\eta+e^{-i\t}\eta^c \right) &=0\\
\underline{F}\gamma_m\left(e^{i\t}\eta+e^{-i\t}\eta^c \right) &=0
~.}}
Note that the RR and NSNS equations have completely decoupled. From the first line of (\ref{susyiian}) it immediately follows that the external manifold is constrained to be Minkowski:
$\mcal_2=\mathbb{R}^{1,1}$.
Finally using \eqref{toreta} the Killing spinor equations above reduce to the following set of algebraic relations:
\eq{\spl{
W &= 0\\
\alpha &= e^{ \frac12 A}\\
h_1 &= - 2\d A~,}}
with $A$ unconstrained\footnote{Integrating the
relation $\d(\log\alpha-\frac12 A)=0$,
we have absorbed without loss of generality the integration constant in the definition of $\alpha$.}, while the magnetic fluxes obey a (twisted) self-duality condition:
\eq{\spl{\label{eq:11fluxes1}
F=\star_8\sigma(F)
~.}}
More explicitly we have: $F_4=\star_8F_4$, $F_6=-\star_8F_2$, $F_8=\star_8F_0$, so in particular the four-flux is self-dual.
In addition, the RR-fluxes obey the following relations:
\eq{\spl{  \label{eq:rrconstraints}
f_4&=\frac16 f_0+\frac43 e^{-i\theta}\cos\theta\tilde{f}_4\\
f_2&=2e^{-i\theta}\sin\theta\tilde{f}_4 \\
\sin \t f_{2|mn}^{(2,0)} &= - \cos \t f_{4|mn}^{(2,0)} - \frac{1}{8} e^{i \t} \O_{mn}^{\phantom{mn}pq} f_{4|pq}^{(0,2)} \\
\sin \t f_{4|mn}^{(2,0)} &=   \cos \t f_{2|mn}^{(2,0)} - \frac{1}{8} e^{i \t} \O_{mn}^{\phantom{mn}pq} f_{2|pq}^{(0,2)}
\;,}}
where $f_0$, $f_2$, $f_4$ are real scalars while $\tilde{f}_4$ is complex. Note that the last two equations are equivalent for $e^{2i \t} \neq 1$, whereas for $e^{2 i \t} = 1$ they become independent pseudoreality conditions, cf. \eqref{pseudo}. For the remaining NS fields, we distinguish between three cases.\\

$\bullet$ $e^{2i\t} = 1$:
\eq{
\spl{
e^\phi &=  g_s e^A\\
h_{3}^{(1,0)} = \tilde{h}_{3}^{(1,0)} &= 0\\
h_{3}^{(2,1)} &=0\\
W_1 &=-\frac{3i}{4}W_4\\
W_3 &=\frac{1}{2}W_2\\
W_5 &=\frac{3}{2}W_4
~,}}
with $g_s$ a non-zero integration constant, $A$, $W_4^{(1,0)}$, $W_2^{(2,1)}$ unconstrained.\\

$\bullet$ $e^{2i\t} = -1$:
\eq{
\spl{
h_{3}^{(1,0)} &= 0\\
\tilde{h}_{3}^{(1,0)} &=  \frac{1}{4}\partial^{+}(A-\phi)\\
W_1 &=0\\
W_{2} &=-2i h_{3}^{(2,1)}\\
W_3 &=0\\
W_4 &=\partial^{+}(\phi - A)\\
W_5 &=\frac{3}{2}\partial^{+}(\phi - A)
~,}}
with $A$, $\phi$,  $h_3^{(2,1)}$ unconstrained; for any scalar $S$, $\partial^{\pm}S$ denotes the projection of the exterior derivative $\d S$ onto its (1,0), (0,1) parts.\\

$\bullet$ $e^{2i\t} \neq \pm 1$:
\eq{
\spl{
e^\phi&=g_s e^{A}\cos\theta \\
h_{3}^{(1,0)} &= \frac{2}{3}\partial^{+}\t\\
\tilde{h}_{3}^{(1,0)} &= \frac{1}{4}(i+\tan\t)\partial^{+}\t\\
W_1^{(1,0)}&=\frac{1}{4}(1+i\cot\t)\partial^{+}\t\\
W_{2}^{(2,1)} &=2(-i+\cot\t)~\!h_{3}^{(2,1)}\\
W_3^{(2,1)}&=\cot\t ~\!h_{3}^{(2,1)}\\
W_4^{(1,0)}&=-(\tan\t+\frac13\cot\t)\partial^{+}\t\\
W_5^{(1,0)}&=(i-\frac{1}{2}\cot\t-\frac{3}{2}\tan\t)\partial^{+}\t
~,}}
with $g_s$ a non-zero integration constant,  $A$, $\t$,  $h_3^{(2,1)}$ unconstrained.

\subsection{Integrability for low-dimensional compactifications}
\label{sec:int}

For bosonic type II supergravity backgrounds it is well-known that under certain mild assumptions the supersymmetry of the background together with the (generalized) Bianchi identities for all fields guarantees that the equations of motion of all NSNS fields (i.e. all remaining equations of motion) are automatically satisfied \cite{lt,papiib,gaunteqs,kt,march}. However, for backgrounds of the form $\mathbb{R}^{1,d-1}\times\mcal_{10-d}$ with $d\leq 2$ the assumptions of this integrability theorem are violated, in general. Since this subtlety may not be widely known, it will be useful to clarify this point here following the demonstration in \cite{march}, specializing to the case without sources.

Consider an arbitrary supersymmetric, bosonic type IIA/B supergravity background. Let us define the following tensors:
\beal
E_{MN}&\equiv R_{MN}+2\nabla_{M}\nabla_N\phi-\frac12H_M\cdot
H_N-\frac14e^{2\phi}F^{\mathrm{tot}}_M\cdot F^{\mathrm{tot}}_N\nn\\
\delta H &\equiv e^{2\phi}*_{10}\left[
\d(e^{-2\phi}*_{10} H) -\frac12(*_{10} F^{\mathrm{tot}}\wedge F^{\mathrm{tot}})_{8}
\right]\nn\\
D&\equiv 2R-H^2+8\left(\nabla^2\phi-(\partial\phi)^2\right)~,\label{dlth}
\end{align}
so that the system
\eq{\label{nseqs}E_{MN}=0~;~~~\delta H_{MN}=0~;~~~D=0~,}
is equivalent to the  equations of motion for the metric, the NSNS three-form  and the dilaton. Let us furthermore assume that the (generalized) Bianchi identities\footnote{We work in the democratic
formalism of type II supergravity. This means in particular
that the generalized Bianchi identities for the RR fields, $\d_HF^{\mathrm{tot}}=0$,
are equivalent in the `ordinary' formulation of type II supergravity to both Bianchi identities and equations of motion for the RR fields.} for all fields,
\eq{\label{genids}\d H=0~;~~~\d_HF^{\mathrm{tot}}=0~,}
are satisfied in the background.
A lengthy but straightforward calculation then shows that the supersymmetry of the background implies (cf. (11.4) of \cite{march}):
\eq{\label{int}\mathcal{E}^+_{NP}\Gamma^P\epsilon_1=0~;~~~
\mathcal{E}^-_{NP}\Gamma^P\epsilon_2=0~;~~~D=0
~,}
where $\epsilon_1$, $\epsilon_2$ are the $Spin(1,9)$ Majorana-Weyl Killing spinors of the background, and we have defined,
\eq{\label{curlyedef}
\mathcal{E}^{\pm}_{NP}\equiv
-2 E_{NP}\pm \delta H_{NP}
~.}
Let us multiply the first, the second equation of (\ref{int}) on the left by $\mathcal{E}^+_{NK}\Gamma^K$, $\mathcal{E}^-_{NK}\Gamma^K$ respectively. Expanding the gamma-matrix product $\Gamma^K\Gamma^P$ we obtain:
\eq{\label{i1}
\sum_{P=0}^{9}\mathcal{E}^{\pm}_{NP}\mathcal{E}^{\pm}_{N}{}^P=0
~,~~~ \mathrm{no ~sum~ over}~N~.}
Let us assume in addition that  $\mathcal{E}^{\pm}_{NP}=0$, for $P=0$, $N\neq0$. This assumtion is equivalent to $E_{NP}=0$, $\delta H_{NP}=0$, for $P=0$, $N\neq0$;
since $\mathcal{E}^{+}_{NP}=\mathcal{E}^{-}_{PN}$, it is also equivalent to
$\mathcal{E}^{\pm}_{NP}=0$, for $P\neq0$, $N=0$.
Eq.(\ref{i1}) then implies
\eq{\label{i2}
\sum_{P=1}^{9}\mathcal{E}^{\pm}_{NP}\mathcal{E}^{\pm}_{N}{}^P=0
~,~~~ N\neq0~.}
Since for fixed $N$ the equation above can be thought of
as the vanishing of the Euclidean norm of a vector, it follows that
$\mathcal{E}^{\pm}_{NP}=0$, for $N\neq0$ and for all $P$ (remember that $\mathcal{E}^{\pm}_{N0}=0$, for $N\neq0$ by assumption), and hence also
$\mathcal{E}^{\pm}_{NP}=0$, for $P\neq0$ and for all $N$. Taking the
latter equation into account and setting $N=0$ in (\ref{i1}) we thus obtain
$\mathcal{E}^{\pm}_{00}=0$.

The upshot of the analysis of the previous paragraph is that {\em under the assumption of the vanishing of the mixed time-space components: $E_{NP}=0$, $\delta H_{NP}=0$, for $P=0$, $N\neq0$, supersymmetry together with the (generalized) Bianchi identities for all fields imply that all remaining equations of motion (i.e. those for the metric, the dilaton and the NSNS three-form) are automatically satisfied}.

For backgrounds which are topologically of the form $\mathbb{R}^{1,d-1}\times\mcal_{10-d}$ or AdS$_{d}\times\mcal_{10-d}$ with $d>2$, provided that all fields (including the warp factor) only depend on the coordinates of the internal manifold $\mcal_{10-d}$ the assumptions of the integrability theorem  above can be seen to be satisfied. For $d=2$ however it is possible for $\delta H_{NP}$ to have non-vanishing mixed time-space components without violating two-dimensional Lorentz invariance -- since in this case $\delta H$ can be proportional to the volume element of $\mathbb{R}^{1,1}$ or AdS$_{2}$. For $d=1$ both $E_{NP}$ and $\delta H_{NP}$ can have non-vanishing mixed time-space components.

We conclude that for low-dimensional compactifications $d=1,2$ the assumptions of the integrability theorem above are not, in general, automatically satisfied.
For the backgrounds we are considering in this paper, the additional condition that needs to be imposed, besides supersymmetry and the (generalized) Bianchi identities for the fluxes, turns out to be a second-order differential equation for the warp factor $A$, cf. (\ref{int2}), (\ref{u11}) below. All remaining equations of motion are then guaranteed to be automatically satisfied.

\subsection{$\ncal=(1,1)$ Calabi-Yau compactifications}\label{sec:iiath}
Let us now examine the special case where $\mathcal{M}_8$ is Calabi-Yau. From the point of view of $G$-strctures, the natural definition of a Calabi-Yau manifold is a manifold with $SU(n)$-structure such that all torsion classes vanish: in our case we thus consider fourfolds with $SU(4)$-structure satisfying $W_i=0$, $i=1,\dots , 5$. Note that this means that the holonomy group is allowed to be a proper subgroup of $SU(4)$. As a consequence we do not necessarilly have $\mathrm{dim}H^{2,0}(\mcal_8) = 0$, as would be the case if the holonomy where exactly $SU(4)$ \cite{yau}.

Applying the results of section \ref{medium}, we have the following supersymmetric $\ncal=(1,1)$ IIA background. The ten-dimensional metric is given by
\eq{
ds^2=e^{2A}ds^2(\mathbb{R}^{1,1})+ ds^2(\mcal_8)
~,}
with $\mcal_8$ a Calabi-Yau fourfold. The vanishing of the torsion classes means that all branches of the solution of section \ref{medium} can be parameterized as follows: the dilaton and the NSNS
three-form $H$ are given by
\eq{\label{hf}
e^{\phi}=g_se^A~;~~~
H=-\mathrm{vol}_2\wedge\d e^{2A}
~,
}
where $g_s$ is a constant and $\mathrm{vol}_2$ is the unwarped volume element of $\mathbb{R}^{1,1}$, while the magnetic RR forms are given by
\eq{\spl{\label{rrf}
F_0&=f_0~;~~~F_2=f_2 J + f_2^{(1,1)}+ \left(f_2^{(2,0)} + \text{c.c.}\right)\\
F_4&= \star_8F_4 = f_4^{(2,2)} + f_4 J \wedge J + \left( \tilde{f}_4 \O + f^{(2,0)}_4 \wedge J + \text{c.c} \right)
\\
F_6&=-\star_8F_2~;~~~
F_8=\star_8F_0
~,
}}
where $J$ and $\O$ are the K\"{a}hler form and holomorphic four-form of the Calabi-Yau fourfold respectively; $f_0$, $f_2$, $f_4$ are real scalars while $\tilde{f}_4$ is complex, $f_2^{(2,0)}$, $f_4^{(2,0)}$ are (2,0) two-forms, $f_2^{(1,1)}$ is a primitive (1,1) two-form and $f_4^{(2,2)}$ is a primitive (2,2) four-form. Moreover the RR fluxes obey the constraint \eqref{eq:rrconstraints} with $\t$ now constrained to be constant.

For this background, the Bianchi identity \eqref{genids} for $H$ is automatically satisfied, while
the (generalized) Bianchi identites for the RR fields $\d_H F^{\mathrm{tot}}=0$ reduce to the condition that the magnetic RR flux $F$ should be closed and co-closed,
\eq{\d F=\d\star_8 F=0~.}
More explictily this means that $f_0$, $f_2$, $f_4$, $\tilde{f}_4$ are constant while $f_2^{(1,1)}$, $f_2^{(2,0)}$, $f_4^{(2,0)}$, $f_4^{(2,2)}$ are harmonic.\footnote{A closed and co-closed form is necessarily harmonic. Conversely on a smooth, compact Riemannian manifold harmonic forms are closed and co-closed} In addition, according to the integrability theorem, we must impose the vanishing of the mixed time-space components of $\delta H_{MN}$. Substituting the profiles of $H$, $F$ given above, this condition can be seen to be equivalent to:
\eq{\label{int2}
-\d\star_8\d e^{-2A}+\frac{g_s^2}{2} F\wedge \sigma(F)|_8=0
~.
}
The integrability theorem guarantees then that all remaining equations of motion are automatically satisfied. The equation above can be generalized to include a number of spacetime-filling fundamental strings.

Note that the integral over $\mcal_8$ of the second term on the left-hand side of (\ref{int2}) is non-negative,
\eq{
\frac12 \int_{\mcal_8}F\wedge \sigma(F)
=\int_{\mcal_8}\left(F_0\wedge\star F_0+F_2\wedge\star F_2+\frac12 F_4\wedge\star F_4\right)\geq 0~,}
thanks to the twisted self-duality of the magnetic flux, cf. (\ref{rrf}).
Assuming $\mcal_8$ is smooth and compact, integrating (\ref{int2}) over $\mcal_8$ thus implies that all flux should vanish and the warp factor $A$ must be  constant. This is of course a special case of the no-go theorem of \cite{mn}.
The no-go theorem can be evaded by taking $\mcal_8$ non-compact or singular.
 It can also be evaded by
including higher-order corrections  -- we will discuss such corrections in more detail in section \ref{sec:mth}.

\section{Supersymmetric M-theory backgrounds with $SU(4)$ structure}\label{sec:mth}

By setting to zero the Romans' mass,
\eq{\label{u1} f_0=0~,}
the 2d, $\ncal=(1,1)$ solution of section \ref{medium} can be uplifted to a 3d,  $\ncal=1$ (two real supercharges) solution of eleven-dimensional supergravity on eight-manifolds with  $SU(4)$ structure.
As can be deduced from appendix \ref{app:uplift},
in order for the ten-dimensional metric (\ref{10dmetric})
to uplift to a warped product of the form $\mathbb{R}^{1,2}\times\mcal_8$ we must set $\phi=A+\phi_0$, where $\phi_0$ is a constant which may be set to zero without loss of generality by rescaling the coordinates. In addition we must have $C_1=0$, where $C_1$ is the potential for the RR two-form $F_2=\d C_1$.
This requires setting
\eq{\label{u3}
f^{(1,1)}_{2|mn}=f^{(2,0)}_{2|mn}=0~;~~~f_2=0
~,}
in the solution  of section \ref{medium}. We may also
choose the branch of the solution with $e^{2i\theta}=1$, as
the other branches turn out to be subcases thereof.
We thus obtain the following
expressions for the eleven-dimensional metric,
\eq{\label{u2}
ds^2=e^{\frac{4A}{3}}ds^2(\mathbb{R}^{1,2})+e^{-\frac{2A}{3}}ds^2(\mcal_8)
~,}
and the eleven-dimensional four-form,
\eq{\label{u4}
G=\mathrm{vol}_3\wedge\d(e^{2A})+
f_4 \big(J \wedge J +\frac{3}{2}~\!\mathrm{Re}\O \big) + (f_4^{(2,0)} - \frac{1}{4} f_4^{(2,0)} \lrcorner \O^* ) \wedge J
+f_4^{(2,2)}~,
}
where $\mathrm{vol}_3$ is the unwarped volume element of $\mathbb{R}^{1,2}$, $f_4$ is an unconstrained real scalar, and
$f_4^{(2,2)}$ is the (2,2) primitive part of the four-form flux; our conventions for the contraction of forms are given in footnote \ref{f16}. We have also taken into account that the (2,0)-form $f_4^{(2,0)}$ obeys the pseudoreality
condition
\eq{\label{35}
f_4^{(2,0)*} = - \frac{1}{4} f_4^{(2,0)}\lrcorner \O^*
~,}
as follows from (\ref{eq:rrconstraints}).
The internal part of the four-form is self-dual, as follows from (\ref{dec4s}), (\ref{dec4a}). However, in contrast to the $\ncal=2$ solutions of \cite{beck}, it need not be primitive nor does its (4,0) part need to vanish. Rather we see that a certain combination
of non-primitive flux and (4,0) flux  (the component in (\ref{u4})
which is proportional to $f_4$) can preserve $\ncal=1$ supersymmetry. Additionally, there is a non-primitve $(3,1)$ term.
Moreover, the torsion classes of $\mcal_8$ are given by:
\eq{\label{u5}
W_3=\frac{1}{2}W_2~;~~~W_4=\frac{4i}{3}W_1~;~~~W_5=2i W_1~,
}
with $W_1$, $W_2$ unconstrained. Again we see that in contrast to \cite{beck} the torsion classes need not vanish; i.e. the internal manifold need not be a Calabi-Yau fourfold.

The previous analysis can be verified directly in eleven dimensions: Let us make the following ansatz for the eleven-dimensional metric and four-form:
\eq{\spl{ds^2&=e^{\frac{4A}{3}}\left(
ds^2(\mathbb{R}^{1,2})+\widetilde{\d s}^2(\mcal_8)
\right)\\
G&=
F+e^{2A}\mathrm{vol}_3\wedge f
~,}}
where the metric $\tilde{g}_{mn}$ on $\mcal_8$ above is related to
the metric ${g}_{mn}$ in (\ref{u2}) through a Weyl rescaling:
\eq{\label{u6}
\tilde{g}_{mn}=e^{-2A}{g}_{mn}
~.}
Let us furthermore express the Killing spinor $\epsilon$ of
eleven-dimensional supergravity as
\eq{\label{u7}
\epsilon=e^{-\frac{A}{3}}\zeta\otimes(\eta+\eta^c)
~,}
where $\zeta$ is a Majorana spinor of $Spin(1,2)$. The Killing spinor above  parametrizes a 3d, $\ncal=1$ supersymmetry ansatz, i.e. two real supercharges -- as many as the independent components of $\zeta$. In the conventions of \cite{mb,tsimpism} it then follows that the supersymmetry equations are equivalent to the following set of conditions:\footnote{We have set the warp factor $\Delta$ of \cite{mb,tsimpism} equal to $2A/3$ here.
Equations (\ref{u8}) are obtained from (3.5)-(3.8) of \cite{tsimpism} by setting $\xi^+=\eta+\eta^c$, $m=\xi^-=0$ therein together with   $\widetilde{\nabla}\xi^+=\nabla_m\xi^+-\frac{1}{2}\gamma_m{}^n\partial_n A$ which follows from (\ref{u6}); $\widetilde{\nabla}$ is the covariant derivative with respect to the
Christoffel connection of $\tilde{g}_{mn}$.}
\eq{\spl{\label{u8}
{\nabla}_m\eta+\mathrm{c.c.}&=0\\
2\partial_mA-f_m&=0\\
F_{mpqr}\gamma^{pqr}\eta+\mathrm{c.c.}&=0
~,}}
where $\nabla_m$ is the covariant derivative with respect to the
Christoffel connection of ${g}_{mn}$. Using (\ref{toreta}), (\ref{c4}), (\ref{c7}), the system above can be seen to be equivalent to (\ref{u4}), (\ref{u5}), with the eleven-dimensional metric given by (\ref{u2}).

In addition to the supersymmetry equations we have the following Bianchi identities and equations of motion:
\eq{\spl{\label{u11}
0&=\d F=\d\star_8 F\\
0&=-\d\star_8\d e^{-2A}+\frac12 F\wedge F
~,}}
where we have taken into account the self-duality of $F$ and the
relation $f=2\d A$, following from the supersymmetry equations (\ref{u8}).
 As can be seen e.g. by uplifting the integrability theorem of section \ref{sec:int}, cf. (\ref{int2}), or directly in eleven dimensions \cite{inta}, the supersymmetry equations (\ref{u8}) together with (\ref{u11}) imply all remaining equations of motion.

Note that imposing instead a 3d $\ncal=2$ supersymmetry ansatz of the form
\eq{\label{u9}
\epsilon=e^{-\frac{A}{3}}(\zeta\otimes\eta+\mathrm{c.c.})
~,}
where now $\zeta$ is a complexified spinor of $Spin(1,2)$,
results in the stronger system
\eq{\spl{\label{u}
{\nabla}_m\eta&=0\\
2\partial_mA-f_m&=0\\
F_{mpqr}\gamma^{pqr}\eta&=0
~,}}
which implies that $\mcal_8$ is a Calabi-Yau fourfold (all torsion classes vanish) and the four-form flux is given by
\eq{\label{u10}
G=\mathrm{vol}_3\wedge\d(e^{2A})
+f_4^{(2,2)}
~.
}
In particular the internal part of the flux is (2,2) and primitive.
We thus recover the case analyzed in \cite{beck}.

In the absence of negative-tension sources or higher-order derivative corrections, provided $\mcal_8$ is a smooth compact manifold, the $\ncal=1$ backgrounds of the form $\mathbb{R}^{1,2}\times\mcal_8$ we have been considering run into the no-go theorem of
\cite{mn}: the equations of motion of the fields imply that all flux vanishes and $\mcal_8$ is Ricci-flat. This conclusion can be avoided  by including  higher-order derivative corrections to the equations of motion.
More specifically, as was first noted in \cite{beck}, the equation of motion for the warp factor, i.e. the second line in (\ref{u11}), receives a higher-order correction related to the fivebrane anomaly \cite{dlm}:\footnote{In our units the M2 brane tension $T_{\mathrm{M2}}=\frac{2\pi}{l_P^3}$ is set to one, with $l_P$ the eleven-dimensional Planck length.}
\beal
-\d\star_8\d e^{-2A}+\frac12 F\wedge F+(2\pi)^2\sum_{i=1}^{N_{\mathrm{M2}}}\delta^{(8)}(y-y_i)\mathrm{vol}_8=(2\pi)^2 X_8~,
\label{x8}
\end{align}
where $X_8$ can be expressed in terms of the first and second Pontrjagin
forms $p_1$, $p_2$:
\eq{\label{x8def}
X_8 =\frac{1}{48}~\!(p_2-\frac14 p_1^2)
~,}
and on the left-hand side of (\ref{x8}) we have added $N_{\mathrm{M2}}$ sources of M2 branes which are localized at the points $y_i\in\mcal_8$, $i=1,\dots,N_{\mathrm{M2}}$, and are extended along $\mathbb{R}^{1,2}$; $\mathrm{vol}_8$ is the unwarped volume element of $\mcal_8$.
The integrated version
of equation (\ref{x8}) thus provides a topological constraint \cite{svw,dm},
\beal\label{tadpole}
\frac{1}{8\pi^2} \int_{\mcal_8}
F\wedge F+N_{\mathrm{M2}}=\frac{\chi(\mcal_8)}{24}~,
\end{align}
where $\chi(\mcal_8)$ is the Euler characteristic of $\mcal_8$, cf. (\ref{top}). This constraint is the integrability condition for the existence of a solution to the differential equation (\ref{x8}).

There are other corrections at the same order in the derivative expansion
of eleven-dimensional supergravity, notably corrections to the
supersymmetry variations, which should also be taken into account, see e.g. \cite{pvwa, cetal, ht}. In general it is inconsistent to consider only the correction (\ref{x8}) above  without at the same time modifying the supersymmetry variations. However it can be argued, see e.g. \cite{pvw}, that
in the large-volume approximation it is consistent
to take only the above correction into account. We will consider the
large-volume approximation in more detail in section \ref{sec:cy}.

\subsection{$\mathcal{N}=1$ Calabi-Yau compactifications}\label{sec:cy}
Similarly to what we did for IIA in section \ref{sec:iiath}, we will now specialize to the case where $\mcal_8$ is Calabi-Yau. Minimally supersymmetric Calabi-Yau compactifications are obtained from the results of the previous section by setting to zero all the torsion classes in (\ref{u5}). We will restrict ourselves to the case where the holonomy of $\mcal_8$ is exactly $SU(4)$ and not a proper subgroup thereof.
Let us collect the results here for the convenience of the reader.

The metric is of the form of a
warped product,
\eq{\label{423}ds^2=e^{\frac{4}{3}A}ds^2(\mathbb{R}^{1,2})+e^{- \frac{2}{3} A} ds^2(\mcal_8)
~,}
where now $\mcal_8$ is a Calabi-Yau fourfold.
The four-flux
is given by
\eq{\label{43}G=
\mathrm{vol}_3\wedge\d(e^{2A})+F~;~~~F=
f_4 \left(J \wedge J +\frac{3}{2}~\!\mathrm{Re}\O \right)
+f_4^{(2,2)}~.}
Its internal part $F$ is constrained to be harmonic,
\eq{\label{44}\d F=\d\star_8 F=0~,}
hence $f_4$ is a (real) constant, while $f_4^{(2,2)}$ is harmonic.
Note that, in comparison with \eqref{u4}, $f_4^{(2,0)}$ has been set to zero. This is a consequence of the fact that for $\mcal_8$ a manifold of strict $SU(4)$ holonomy there are no non-trivial harmonic (2,0) two-forms \cite{yau}.
The warp factor satisfies the differential equation
\beal
-\d\star_8\d e^{-2A}+\frac12 F\wedge F+(2\pi)^2\sum_{i=1}^{N_{\mathrm{M2}}}\delta^{(8)}(y-y_i)\mathrm{vol}_8=(2\pi)^2 X_8~.
\label{x82}
\end{align}
The existence requirement for a solution to the equation above is given by  the topological constraint:
\beal\label{tadpole2}
\frac{1}{8\pi^2} \int_{\mcal_8}
F\wedge F+N_{\mathrm{M2}}=\frac{\chi(\mcal_8)}{24}~.
\end{align}
All remaining equations of motion are automatically satisfied.

Furthermore in the quantum theory the four-form flux $F$ is also subject to the quantization condition \cite{wq}:
\eq{\label{wqc}
\frac{[F]}{2\pi}-\frac{p_1}{4}\in H^4(\mcal_8,\mathbb{Z})
~,}
where $[F]$ is the cohomology class of $F$. For future use, let us also mention that for $S$ a $\mathrm{K3}$ surface,
\eq{
-\int_{S}\frac{p_1}{2}
=\int_S{c_2}
=\chi(S)=24
~,}
where in the first equality we have used that $p_1=c_1^2-2c_2$ and that
$c_1=0$ for a K3 surface. Specializing to the case $\mcal_8=S\times\tilde{S}$ with $S$, $\tilde{S}$ K3 surfaces, the quantization condition (\ref{wqc}) can therefore be written equivalently as:
\eq{\label{qk3}
\frac{1}{2\pi}\int_{\mathcal{C}_4}F\in\mathbb{Z}
~,}
for any homology cycle $\mathcal{C}_4$ such that $\{\mathcal{C}_4\}$ forms a
basis of $H_4(S\times\tilde{S},\mathbb{R})$ dual to the basis of $H^4(S\times\tilde{S},\mathbb{Z})$; in the case at hand we may take $\mathcal{C}_4$ in (\ref{qk3}) to be any element of the integral homology lattice $H_4(S\times\tilde{S},\mathbb{Z})$.

Let us now compare the general solution above with the large-volume approximation for $\mcal_8$ \cite{ww}. In the present context the analysis has been carried out in \cite{beckersusybr}, specializing to the case without M2 branes $N_{\mathrm{M2}}=0$,\footnote{The large-volume analysis of \cite{beckersusybr} can be  generalized to include a non-zero number of localized M2 branes. In that case the warp factor is no longer constant at lowest order, but rather satisfies (\ref{x82}) with $N_{\mathrm{M2}}\neq0$. Correspondingly at lowest order the metric  is a warped product and the four-form flux is not purely internal, in accordance with (\ref{423}), (\ref{43}).} and takes into account certain higher-order corrections to eleven-dimensional supergravity at eighth order in derivatives.
The large-volume limit consists in expanding
\eq{
g_{mn}=t^2g^{(0)}_{mn}+g^{(2)}_{mn}+\dots~,
}
where $t\rightarrow\infty$ is the radius\footnote{It is perhaps more  accurate to say that
we are considering the limit $t>>l_P$, where $l_P$ is the Planck length and $t^8=\mathrm{vol}(\mcal_8)$.} of $\mcal_8$
and similarly for the other supergravity fields, and analyzing the
equations of motion at each order in $t$. To leading order in $t$ the result of this analysis is that \cite{beckersusybr}:
\begin{itemize}
\item The internal manifold $\mcal_8$ is Ricci-flat.
\item The warp factor $A$ is constant.
\item The four-form flux is is purely internal; it is closed, self-dual (hence harmonic) and satisfies  (\ref{tadpole2}) with $N_{\mathrm{M2}}=0$.
\end{itemize}
Note that this is for arbitrary vacua, supersymmetric or not.
Comparison with our general analysis of supersymmetric Calabi-Yau vacua leads to the conclusion that the first condition together with the self-duality and harmonicity of the internal part $F$ of the four-flux are  automatically satisfied, whereas the remaining conditions are additional constraints imposed by the large-volume limit and the assumption of absence of M2 branes.

\subsubsection{Superpotentials}\label{sec:supo}

The conditions for $\ncal=1$ supersymmetry of the vacua of section \ref{sec:cy}  are equivalent to those of \cite{haacksamt}, where the conditions for a partial $\ncal=1$ supersymmetry breaking in three-dimensional $\ncal=2$  gauged supergravity were derived for the first time.\footnote{The analysis of
\cite{haacksamt} was performed in the $N_{\mathrm{M2}}=0$, large-volume limit. The Calabi-Yau vacua of section \ref{sec:cy} are more general in that they allow for a nontrivial warp factor and a nonvanishing number of spacetime-filling M2 branes.} Our presentation has been slightly different from that of \cite{haacksamt}. Recall that in deriving the general form of the four-form flux (\ref{43}) we have ``gauge-fixed''  the covariantly constant  spinor $\eta$ of the Calabi-Yau fourfold by absorbing a phase in the definition of $\eta$ so that $\alpha$ in (\ref{spindecompIIA}) is real. Reinstating an arbitrary constant phase $\eta\rightarrow e^{\frac{i}{2}\vartheta}\eta$ results in $\O\rightarrow e^{i\vartheta}\O$, as follows from (\ref{b3}), hence the expression for the internal part of the flux (\ref{43}) becomes:
\eq{\label{f}F=f_4^{(2,2)}+
f_4 J \wedge J +\left(\tilde{f}_4\O +\mathrm{c.c.}\right)~,}
where we have set $\tilde{f}_4\equiv\frac{3}{4}f_4e^{i\vartheta}$, so that $|\tilde{f}_4|=\frac{3}{4}f_4$. This is precisely as in \cite{haacksamt}, cf. section 3.3 therein.
In other words, the large-volume vacua of section \ref{sec:cy} can be thought of as the uplift to eleven dimensions of partial $\ncal=1$ supersymmetry-breaking vacua of three-dimensional $\ncal=2$  gauged supergravity.
Since compactification of M-theory on manifolds of $Spin(7)$ holonomy yields an effective three-dimensional $\ncal=1$ supergravity and since Calabi-Yau fourfolds are special cases of $Spin(7)$ manifolds, we expect that these vacua can also be thought of as $\ncal=1$ vacua of three-dimensional $\ncal=1$
supergravity. As we will see this expectation is borne out by our results.\footnote{It is interesting to note the analogy with the four-dimensional $\ncal=2$ vacua of type II string theory on
Calabi-Yau threefolds with vanishing Euler characteristic, recently considered in \cite{kmt}: since these threefolds admit an $SU(2)$ structure and since compactification of type II supergravity on six-dimensional manifolds with $SU(2)$ structure results in an effective four-dimensional
$\ncal=4$ supergravity, these vacua can be thought of as spontaneously broken $\ncal=2$ vacua of $\ncal=4$ supergravity. At the same time they can be thought of as $\ncal=2$ vacua of the standard $\ncal=2$ supergravity obtained by compactification of type II on Calabi-Yau threefolds.}

Before examining the  $\ncal=1$  case let us recall the superpotential conditions in the case of  $\ncal=2$ compactifications \cite{gvw,hl} (see also \cite{gukov,hg} for the case of type IIA compactifications on Calabi-Yau fourfolds).
These conditions can be derived from the two superpotentials
\eq{W=\int_{\mcal_8} F \wedge \Omega,}
and
\eq{{\hat W}={1\over 2} \int_{\mcal_8}F\wedge J \wedge J~,}
where $\Omega$ and $J$ are the holomorphic four-form and the K\"ahler
form of the Calabi-Yau manifold $\mcal_8$ respectively.
Note that although ${\hat W}$ is not a holomorphic function (of the complexified K\"{a}hler moduli) it is usually still called a superpotential in the literature.
The conditions for $\ncal=2$
compactifications to three-dimensional Minkowski space are:
\eq{\label{n2sp}W={\hat W}=0~;~~~
D_{\alpha}W={D}_A{\hat W}=0~,}
where $\alpha=1,\dots,h^{1,3}$
parameterize the (complex) deformation moduli of the complex structure
whereas $A=1,\dots,h^{1,1}$ parametrize the (real) deformation moduli of the K\"ahler structure. The K\"{a}hler-covariant derivatives are defined as:
\eq{
D_{\alpha}W={\partial}_{\alpha}W-({\partial}_{\alpha}K)W~;~~~
D_{A}\hat W={\partial}_A {\hat W}-{1\over 2} ({\partial}_AK){\hat W}
~,
}
where $K=K_{\Omega}+K_J$,
\eq{K_{\Omega}=\log \int_{\mcal_8}\frac{1}{16}\Omega \wedge \Omega^*~;~~~K_J=\log \int_{\mcal_8}\frac{1}{4!}J^4
~,}
is the  K\"ahler potential of three-dimensional supergravity with four real supercharges.
Let us also decompose the four-form flux in $SU(4)$ modules:
\eq{\label{gdec}
G=
f_4^{(2,2)}+f_4^{(1,1)}\wedge J+f_4J\wedge J
+\big(f_4^{(3,1)}+f_4^{(2,0)}\wedge J+\tilde{f}_4~\!\Omega
+\mathrm{c.c.}\big)
~,}
where we have used exactly the same notation as in (\ref{dec4s}), (\ref{dec4a}) to label the irreducible representations.
With these definitions it follows that
\eq{\spl{\label{ee}
W=16~\!\tilde{f}_4^*~\!\mathrm{vol}(\mcal_8)~&;~~~
\hat W=12~\!f_4~\!\mathrm{vol}(\mcal_8)\\
D_{\alpha}W=\int_{\mcal_8} \big(f_4^{(1,3)}+f_4^{(0,2)}\wedge J\big) \wedge{e}_{\a}  ~&;~~~
D_{A}\hat W=\int_{\mcal_8} f_4^{(1,1)}\wedge J^2  \wedge {E}_{A}
~,
}}
where ${e}_{\a}$ is a basis of $H^{3,1}(\mcal_8)$ while
${E}_{A}$ is a basis of $H^{1,1}(\mcal_8)$; in the first line above
we have taken (\ref{b4}) into account.
The conditions (\ref{n2sp}) can then be seen to
be equivalent in the large-volume limit to the $\ncal=2$ supersymmetry conditions for M-theory on eight-dimensional manifolds found in \cite{beck}.
Indeed, using (\ref{ee}), the first of (\ref{n2sp}) implies $f_4=\tilde{f}_4=0$ while the second gives  $f_4^{(3,1)} = f_4^{(2,0)} = f_4^{(1,1)}=0$. It then follows from (\ref{gdec}) that $G=f_4^{(2,2)}$, i.e. the four-form flux is (2,2) and primitive.

It was shown in \cite{gukovSpin7,BeckerConst} that the  supersymmetry conditions
for compactification on $Spin(7)$-holonomy manifolds can be expressed in terms of a real superpotential, as expected from three-dimensional $\ncal=1$ supergravity \cite{henninga,henningb}.
 Since $SU(4)$ holonomy is a special case of $Spin(7)$ we expect to be able to express the conditions for the $\ncal=1$ Calabi-Yau vacua in terms of a real superpotential. This is indeed possible as follows. First let us reinstate the arbitrary constant phase in the definition of $\Omega$, cf. the discussion around (\ref{f}). A short calculation gives
\eq{\spl{\label{yui}
e^{\frac{1}{2} K_{\Omega}}&\partial_{R^{\alpha}}\left(e^{-\frac{1}{2} K_{\Omega}}|W|\right)=
\mathrm{Re}\left(e^{i\theta}\int_{\mcal_8} \big(f_4^{(1,3)}+f_4^{(0,2)}\wedge J\big) \wedge{e}_{\a} \right) \\
e^{\frac{1}{2} K_{J}}&\partial_{A}\left(e^{-\frac{1}{2} K_{J}}\hat W\right)=\int_{\mcal_8} f_4^{(1,1)}\wedge J^2  \wedge {E}_{A}
~,
}}
where $\partial_{R^{\alpha}}=\partial_{\alpha}+\partial^*_{\alpha}$ and $\tilde{f}_4=|\tilde{f}_4|e^{i\vartheta}$.
The conditions for $\ncal=1$ M-theory compactifications on Calabi-Yau fourfolds given in section \ref{sec:cy} can be written as
\eq{\label{eee}
|W|-\hat{W}=0~;~~~ \partial_{R^{\alpha}}\left(e^{-\frac{1}{2} K_{\Omega}}|W|\right)=0~; ~~~  \partial_{A}\left(e^{-\frac{1}{2} K_{J}}\hat W\right)=0
~.}
Indeed, using (\ref{ee}), (\ref{yui}), the first of (\ref{eee}) is equivalent to  $|\tilde{f}_4|=3/4f_4$; the second gives  $f_4^{(3,1)} =  f_4^{(2,0)}=0$, while the  last one gives $f_4^{(1,1)}=0$. It then follows from (\ref{gdec}) that the four-form flux is as in (\ref{f}). We can now make contact with the formalism of $\ncal=1$ vacua of $\ncal=1$ supergravity in terms of a real superpotential $\mathcal{W}$ by setting:
\eq{\mathcal{W}=
e^{-\frac{1}{2} K_{\Omega}}|W|-e^{-\frac{1}{2} K_{J}}\hat W
~.}
The $\ncal=1$ vacua given in (\ref{eee}) can then be equivalently
expressed as follows:
\eq{\label{ert}\partial_{R^{\alpha}}\mathcal{W}=\partial_{A}\mathcal{W}=\mathcal{W}=0~,}
where we have taken into account that at any point in the moduli space of metrics $K_J$ and $K_{\Omega}$ are both equal to  $\log\mathrm{vol}(\mcal_8)$, cf. (\ref{b4}).\footnote{Thanks to Yau's theorem, a Ricci-flat metric on a Calabi-Yau fourfold $Y_4$ is uniquely determined by specifying a complex
structure and a K\"{a}hler class. Moreover complex structure and K\"{a}hler deformations are independent of each other, which implies that at least locally the moduli space $\mathfrak{M}$ of Ricci flat metrics on $Y_4$ has a direct-product structure, $\mathfrak{M}=\mathfrak{M}_c\times\mathfrak{M}_k$, where $\mathfrak{M}_c$ and $\mathfrak{M}_k$ are the complex structure
and K\"{a}hler moduli spaces repsectively, see e.g. \cite{hthes} and references therein. A choice of $\Omega$, $J$ thus specifies a point in $\mathfrak{M}_c$, $\mathfrak{M}_k$ respectively, and hence a point $\mathfrak{M}$. Viewing $\mathfrak{M}$ as a (trivial) fibration of $\mathfrak{M}_c$ over
$\mathfrak{M}_k$, over each point in $\mathfrak{M}_k$ (specified by some fixed $J$) we have a fiber isomorphic to $\mathfrak{M}_c$
-- whose points correspond to different choices of $\Omega$. Furthermore,
for any $c\in\mathbb{C}^*$, $\Omega$ and $c~\!\Omega$ correspond to the same complex structure on $Y_4$, i.e. the same point in $\mathfrak{M}_c$. Given the fact that the norm of $\Omega$ is constant over $Y_4$, we are thus free to normalize $\Omega$ so that (\ref{b4}) holds for the entire fiber $\mathfrak{M}_c$ over any given point in $\mathfrak{M}_k$.} The above conditions are indeed of the form of \cite{gukovSpin7,BeckerConst} for a $\ncal=1$ supersymmetric Minkowski vacuum of three-dimensional $\ncal=1$ supergravity. Note however that
since $\mathcal{W}=0$ at the vacuum, the real superpotential is only determined up to a (possibly field-dependent) overall normalization. The latter can be fixed by different methods, e.g.  by dimensional reduction of the M-theory action on Calabi-Yau fourfolds, as in \cite{grimm}.

Finally, let us note that the superpotential conditions (\ref{ert}) leave  the volume modulus of $\mcal_8$ unfixed, as they are invariant under an overall volume rescaling. This is compatible with the assumption that we can take the large-volume limit $t\rightarrow\infty$.

\subsection{$\mathcal{N}=1$ compactifications on K3$\times$K3 }\label{sec:k3}
In order to give some simple explicit examples of supersymmetric vacuua
we will now specialize to the case where $\mcal_8$ is a direct product $S\times\tilde{S}$, with $S$, $\tilde{S}$  K3 surfaces. As a consequence there is a further reduction from $SU(4)$ to $SU(2) \times SU(2)$ holonomy.\footnote{In section \ref{sec:cy} we imposed strict $SU(4)$ holonomy in order to make contact with three-dimensional supergravity. For $\mcal_8$ of the form K3$\times$K3 the four-flux may include an additional non-primitive (3,1) term  $f_4^{(2,0)}\wedge J$. We will come back to this point at the end of the section.}
In particular the covariantly constant spinor $\eta$ of $\mcal_8$ decomposes as
\eq{
\eta = \eta_S \otimes \eta_{\tilde{S}}\;,
}
where $\eta_S$, $\eta_{\tilde{S}}$ are covariantly constant spinors of $S$, $\tilde{S}$ respectively. Equivalently, the equation above relates the $SU(4)$ structure of $\mcal_8$ to the $SU(2)$ structures of  $S$, $\tilde{S}$, cf. (\ref{k6}) below. For an introduction to K3 surfaces the reader may consult e.g. \cite{latt,aspinwall} and references therein.

The real cohomology of a compact K3 splits as
\eq{\label{52}
H^2(K3,\mathbb{R})=\mathcal{H}^+(K3,\mathbb{R})\oplus \mathcal{H}^-(K3,\mathbb{R})~,
}
where $\mathcal{H}^{+}(K3,\mathbb{R})$, $\mathcal{H}^-(K3,\mathbb{R})$ are  generated by self-dual, anti-self-dual two-forms respectively; their dimensions
as real vector spaces are equal to three, nineteen respectively. Let $\{j^a , ~ a=1,2,3\}$ and
$\{l^{\alpha}, ~ \alpha=1,\dots,19\}$ be corresponding  bases.
The K\"ahler form and holomorphic two-form $j$, $\omega$ of $S$ can be expanded on the basis of $\mathcal{H}^+(S,\mathbb{R})$ and similarly for the K\"ahler form and holomorphic two-form $\tilde{j}$, $\tilde{\omega}$ of $\tilde{S}$. Explicitly we have:
\eq{\label{k2}
j=\rho^aj^a~;~~~\omega=c^aj^a
~,}
with $\rho^a\in\mathbb{R}$, $c^a\in\mathbb{C}$, $a=1,2,3$.
The volume element $\mathrm{vol}_4$ and the
compatibility relation between $j$, $\omega$ are given by:
\eq{\label{k3}\mathrm{vol}_4=\frac12 j\wedge j=\frac14\omega\wedge\omega^*~;~~~ j\wedge\omega=0~,}
and similarly for $\tilde{S}$. We normalize:
\eq{\spl{\label{k4}
\star_4 j^a= j^a~&;~~~ \star_4l^{\alpha}=-l^{\alpha}\\
\frac12 j^a\wedge j^b=\delta^{ab}\mathrm{vol}_4~&;~~~ \frac12 l^{\alpha}\wedge l^{\beta}=-\delta^{\alpha\beta}\mathrm{vol}_4~,
}}
for $a=1,2,3$, $\alpha=1,\dots,19$, and similarly for
$\tilde{j}^a$, $\tilde{l}^{\alpha}$ of $\tilde{S}$. The normalization above together with (\ref{k2}) gives:
\eq{\label{k5}
\rho^ac^a=0~;~~~c^ac^a=0~;~~~\rho^a\rho^a=1~;~~~c^ac^{*a}=2
~,}
and similarly for $\tilde{\rho}^a$, $\tilde{c}^{a}$. The K\"ahler and
holomorphic forms of $\mcal_8$ $J$, $\Omega$ are given in terms of the corresponding forms of $S$, $\tilde{S}$:
\eq{\label{k6}
J=j+\tilde{j}~;~~~ \Omega=\omega\wedge\tilde{\omega}
~.}
The normalization (\ref{k3}) then implies
\eq{
\frac{1}{4!}J^4=\frac{1}{16}\Omega\wedge\Omega^*
=\mathrm{vol}_4\wedge\widetilde{\mathrm{vol}}_4~;~~~J\wedge\Omega=0~.
}

As in \cite{Bena:2012ub}, in order to examine the conditions for supersymmetry we start with the most general expansion for the self-dual four-form flux $G$:
\eq{
G=f^{ab}j^a\wedge \tilde{j}^b + f^{\alpha\beta} l^{\alpha}\wedge \tilde{l}^{\beta}+f(\mathrm{vol}_4+\widetilde{\mathrm{vol}}_4)
~,}
with $f^{ab}$, $f^{\alpha\beta}$, $f\in\mathbb{R}$.
The condition $\mathcal{W}=0$ in \eqref{ert} is equivalent to
\eq{
f+2(\rho^a\tilde{\rho}^b-c^a\tilde{c}^b)f^{ab}=0
~,}
which in particular implies  $c^a\tilde{c}^bf^{ab}\in\mathbb{R}$.
The condition $\partial_{R^\alpha}\mathcal{W} =0$ is equivalent to
\eq{
\rho^a\tilde{c}^bf^{ab}=c^a\tilde{\rho}^bf^{ab} =0
~,}
where we have taken into account that a basis of $H^{3,1}(S\times\tilde{S})$ is given by $\omega\wedge \tilde{j}$, $\omega\wedge \tilde{l}^{\alpha}$, $\tilde{\omega}\wedge j$ and $\tilde{\omega}\wedge l^{\alpha}$.
Finally, the condition $\partial_{A}\mathcal{W}=0$ gives
\eq{
f^{ab}=(\rho^d\tilde{\rho}^ef^{de})\rho^a\tilde{\rho}^b+\hat{f}^{ab}
~,}
where $\hat{f}^{ab}$ satisfies $\rho^a\hat{f}^{ab}=\tilde{\rho}^b\hat{f}^{ab}=0$.
Putting together all the above, the most general expression for $G$ subject to
the supersymmetry conditions can be written as follows:
\eq{\label{k9}
G=Cj\wedge \tilde{j} +A~\!\mathrm{Re}(\omega\wedge \tilde{\omega})+\mathrm{Re}(B\omega^*\wedge \tilde{\omega})
+(4A-2C)(\mathrm{vol}_4+\widetilde{\mathrm{vol}}_4)
+ f^{\alpha\beta} l^{\alpha}\wedge \tilde{l}^{\beta}
~,}
with $A\in\mathbb{R}$, $B\in\mathbb{C}$, $f^{\alpha\beta}\in\mathbb{R}$
unconstrained. Note that the form of $G$ above was derived using the
superpotential formalism, which automatically imposes the vanishing of $f_4^{(2,0)}$, and is thus suitable for the case of strict $SU(4)$ holonomy.
For $\mcal_8$ of the form K3$\times$K3 one can have an additional non-primitive
(3,1) term (plus its complex conjugate) according to (\ref{u4}), (\ref{35}). The most general such term that can be added to $G$ is thus of the form
\al{\label{nonpri}
(f_4^{(2,0)} +f_4^{(0,2)})\wedge J = \a \o\wedge \tilde{j} + \a^* \tilde{\o} \wedge j+\mathrm{c.c.}\;,
}
with $\a\in\mathbb{C}$ unconstrained.

Let us also mention that in the special case  $A=\alpha=0$ the four-flux takes the form:
\eq{\label{l2}
G=C j\wedge \tilde{j} +\mathrm{Re}(B\omega^*\wedge \tilde{\omega})
-2C(\mathrm{vol}_4+\widetilde{\mathrm{vol}}_4)
+ f^{\alpha\beta} l^{\alpha}\wedge \tilde{l}^{\beta}
~.}
It can be checked that in this case $G$ is of type (2,2) and satisfies
$G\wedge J=0$, so that we recover the conditions of \cite{beck} and supersymmetry is enhanced to $\ncal=2$. Supersymmetry enhancement will be further discussed in the following section.

\subsubsection{Supersymmetry enhancement}\label{sec:enhancement}
The $SU(4)$ structure of $\mcal_8$ is specified by a solution to the Killing spinor equations. Conversely, if multiple $SU(4)$ structures exist for the same background, there will be more free parameters in the expression for the Killing spinor. This leads to supersymmetry enhancement. We will examine this more concretely for $\mcal_8 = S \times \tilde{S}$ by looking at symmetries of the four-flux and the metric.  	

For simplicity in the remainder of this section we will set to zero the non-primitive (3,1) term (\ref{nonpri}).
From the form of the four-flux it follows that the K3$\times$K3 vacua (\ref{k9}) have three-dimensional $\ncal=1$ supersymmetry for generic values of $A$, $B$. To see this, let us note that the Ricci-flat metric on $S$ (and similarly for $\tilde{S}$) is invariant under $SO(3)$ rotations of the triplet ($j,\mathrm{Re}\omega,\mathrm{Im}\omega$). These rotations can be thought of as induced by an $SU(2)$ action on the covariantly constant spinor $\eta_S$ of $S$, under which the latter transforms as a doublet. Since every positive-chirality spinor on $S$ can be expanded on the basis of $\eta_S$ and its complex conjugate $\eta^c_S$, it follows that we can express the  $SU(2)$ action on $\eta_S$ as an $SU(2)$ action on the doublet $\eta_i$, $i=1,2$, where we have defined $\eta_1\equiv\eta_S$, $\eta_2\equiv\eta^c_S$. Explicitly, we define the triplet of real three forms:
\eq{
\vec{J}_{mn}\equiv - \frac{i}{2}\vec{\sigma}_{ij}(\widetilde{\eta^c}_i\gamma_{mn}\eta_{j})
~,}
where $\vec{\sigma}$ is a triplet of Pauli matrices, so that $j=J^{3}$, $\mathrm{Re}\omega=J^{2}$, $\mathrm{Im}\omega=J^{1}$.
Then one can show  that under infinitesimal $SU(2)$ transformations of the spinors $\eta_{1,2}$,
\eq{\label{sdoublet}
\delta\eta_{i}=\frac{i}{2}\delta\vec{\theta}\cdot\vec{\sigma}_{ij}\eta_j
~,}
the forms $J^{a}$ transform as a vector of $SO(3)$:
\eq{\label{jvec}
\delta \vec{J}=-\delta\vec{\theta}\times\vec{J}
~.}
The transformations (\ref{sdoublet}), (\ref{jvec}) leave invariant the associated Ricci-flat  metric of $S$. If the four-form flux were also invariant  there would be a potential enhancement of supersymmetry. Indeed the invariance of the background under (\ref{jvec}) would then mean that there is a family of $SU(4)$ structures (\ref{k6}) and hence
a family of Killing spinors (\ref{u7}) corresponding to the {\em same}  supergravity background (given by the metric and the four-flux), thus enhancing the supersymmetry.

To examine how the four-flux transforms, note that the first term on the right-hand side of (\ref{k9}) breaks (\ref{sdoublet}) and the analogous $SU(2)$ action on the doublet ($\eta_{\tilde{S}}, \eta^c_{\tilde{S}}$) to the $U(1)\times U(1)$ subgroup which acts on $(\omega,\tilde{\omega})$ as
\eq{(\omega,\tilde{\omega})\longrightarrow (e^{i\theta}\omega, e^{i\tilde{\theta}}\tilde{\omega})~.}
The last two terms on the right-hand side of (\ref{k9}) are also invariant under this $U(1)\times U(1)$. The term containing $\omega\wedge \tilde{\omega}$ is invariant under the $U(1)$ subgroup given by $\tilde{\theta}=-\theta$, while the term containing $\omega^*\wedge \tilde{\omega}$ is invariant under a different $U(1)$ subgroup given by $\tilde{\theta}=\theta$. The upshot  is that for generic values of $A$, $B$ the four-form flux $G$ transforms non-trivially  and there is no enhancement of supersymmetry. As we have already mentioned,
this conclusion is no longer valid in the case $A=0$. We also can see this directly as follows: The four-form flux in this case takes the form (\ref{l2}), and is invariant under the $U(1)$ given by
\eq{\label{res}(\omega,\tilde{\omega})\longrightarrow e^{i\theta}(\omega, \tilde{\omega})~.}
This transformation acts non-trivially on the $SU(4)$ structure (\ref{k6}) and is induced by the transformation
$\eta\rightarrow  e^{i\theta/2}\eta$. Thus
in this case the  Killing spinor (\ref{u7}) takes  the form:
\eq{\epsilon=e^{-\frac{A}{3}}\zeta\otimes\eta+\mathrm{c.c.}~,}
with $\zeta$ an arbitrary complexified spinor of $Spin(1,2)$, and
there is supersymmetry enhancement to $\ncal=2$.

Note that although for $B=0$ the four-form flux is invariant under  $(\omega,\tilde{\omega})\rightarrow( e^{i\theta}\omega, e^{-i\theta}\tilde{\omega})$,
this transformation acts trivially on the $SU(4)$ structure (\ref{k6}) and the Killing spinor (\ref{u7}). In this case there is no supersymmetry enhancement.

An example of enhanced $\ncal=4$ supersymmetry is obtained by setting
$A=C$, $B=0$. Indeed in this case the four-flux takes the form:
\eq{\label{another}
G=A~\!j\wedge \tilde{j} +A~\!\mathrm{Re}(\omega\wedge \tilde{\omega})+
2A~\!(\mathrm{vol}_4+\widetilde{\mathrm{vol}}_4)
+ f^{\alpha\beta} l^{\alpha}\wedge \tilde{l}^{\beta}
~,}
and is invariant under the $SO(3)$ transformations given by
\eq{\label{res2}\left( \begin{array}{c}
j\\
\mathrm{Re}\omega \\
\mathrm{Im}\omega
\end{array} \right)
\longrightarrow
R_{\hat{n}}(\theta)\cdot\left( \begin{array}{c}
j\\
\mathrm{Re}\omega \\
\mathrm{Im}\omega
\end{array} \right)~;~~~
\left( \begin{array}{c}
\tilde{j}\\
\mathrm{Re}\tilde{\omega} \\
\mathrm{Im}\tilde{\omega}
\end{array} \right)
\longrightarrow
R_{\hat{n}^{\prime}}(\theta)\cdot\left( \begin{array}{c}
\tilde{j}\\
\mathrm{Re}\tilde{\omega} \\
\mathrm{Im}\tilde{\omega}
\end{array} \right)
~,}
where $R_{\hat{n}}(\theta)$ is an $SO(3)$ rotation  by an angle $\theta$ with respect to the axis given by the unit vector $\hat{n}$; we have defined $\hat{n}'\equiv T\cdot\hat{n}$, where $T=\mathrm{diag(1,1,-1)}$. As follows from the discussion around (\ref{jvec}), the transformation (\ref{res2}) is induced by the $SU(2)$ transformation
\eq{\label{res3}\left( \begin{array}{c}
\eta_S\\
\eta_S^c
\end{array} \right)
\longrightarrow
e^{-\frac{i}{2}\theta\hat{n}\cdot\vec{\sigma}}\cdot\left( \begin{array}{c}
\eta_S\\
\eta_S^c
\end{array} \right)~;~~~
\left( \begin{array}{c}
\eta_{\tilde{S}}\\
\eta_{\tilde{S}}^c
\end{array} \right)
\longrightarrow
e^{\frac{i}{2}\theta\hat{n}'\cdot\vec{\sigma}}\cdot\left( \begin{array}{c}
\eta_{\tilde{S}}\\
\eta_{\tilde{S}}^c
\end{array} \right)
~,}
which acts on the covariantly constant spinors $\eta_S$, $\eta_{\tilde{S}}$ of $S$, $\tilde{S}$. The corresponding transformation of $(\eta+\eta^c)$ reads:
\eq{\eta+\eta^c\longrightarrow a~\!\eta_S\otimes\eta_{\tilde{S}}+b ~\!\eta_S\otimes\eta_{\tilde{S}}^c + \mathrm{c.c.}
~,}
with $a$, $b\in\mathbb{C}$  parameterized by $(\hat{n},\theta)$.
Thus
in this case the  Killing spinor (\ref{u7}) takes  the form:
\eq{\epsilon=e^{-\frac{A}{3}}(\zeta_1\otimes\eta_S\otimes\eta_{\tilde{S}}+
\zeta_2\otimes\eta_S\otimes\eta_{\tilde{S}}^c+
\mathrm{c.c.})~,}
with $\zeta_1$, $\zeta_2$ arbitrary complexified spinors of $Spin(1,2)$, so that
supersymmetry is enhanced to $\ncal=4$.

\subsubsection{K3$\times$K3 examples}\label{sec:examples}
We will now give three explicit examples of K3$\times$K3 vacua, with $\ncal=1,2,4$ supersymmetry repsectively, satisfying the quantization conditions and the topological constraint.  Several  K3$\times$K3 vacua with $\ncal\geq 2$ were presented in \cite{drs}.

The intersection form of $S$ is given by
\eq{\label{intf}
I=U\oplus U\oplus U\oplus(-\mathfrak{e}_8)\oplus(-\mathfrak{e}_8)~,
}
where
$$
U=\left( \begin{array}{cc}
0 & 1 \\
1 & 0
\end{array} \right)~,
$$
and $\mathfrak{e}_8$ is the Cartan matrix of $E_8$.
Let $\Gamma$ be the even self-dual lattice
with (3,19) signature and intersection form (\ref{intf}). By
Poincar\'{e} duality, $H_{2}(S,\mathbb{Z})\cong H^{2}(S,\mathbb{Z}) \cong\Gamma$. Let us enumerate the basis elements of $\Gamma$ as follows: $\{e_{1,I},e_{2,I},e_i\}$, $I=1,2,3$, $i=7,\dots,22$, so that,
\eq{e_{1,I}\wedge e_{2,J}=\delta_{IJ}v_0
~;~~~e_{1,I}\wedge e_{1,J}=e_{2,I}\wedge e_{2,J}=0
~;~~~e_i\wedge e_j=I_{ij}v_0~,}
where $I_{ij}$ is the matrix element of the intersection form (\ref{intf}); $v_0\in H^4(S,\mathbb{Z})$ is the unit volume form of $S$, i.e.
\eq{\int_{S} v_0=1~.}
For later convenience let us also define
\eq{\label{eplus}
e_{\pm I}\equiv e_{1,I}\pm e_{2,I}
~,}
for $I=1,2,3$. It follows that $e_{\pm I}$ has norm $\pm2$ with respect
to the intersection form (\ref{intf}), in particular the $e_{+ I}$ are spacelike.
For a K3 surface  we can choose the basis $\{\mathcal{C}_{1,I},\mathcal{C}_{2,I},\mathcal{C}_i\}$
of the homology lattice $H_{2}(S,\mathbb{Z})$ to be dual to the basis of $\Gamma\cong H^{2}(S,\mathbb{Z})$,
\eq{
\int_{\mathcal{C}_{a,I}}e_{b,J}=\delta_{ab}\delta_{IJ}
~;~~~
\int_{\mathcal{C}_i}e_{j}=\delta_{ij}
~,}
for $a,b=1,2$; $I,J=1,2,3$; $i,j=7,\dots,22$.

The three-plane $\Sigma\equiv\mathcal{H}^+(S,\mathbb{R})$
of (\ref{52}) viewed as a subspace of $\Gamma\otimes\mathbb{R}\cong H^2(S,\mathbb{R})$ is spanned by ($j,\omega$) and is spacelike in the sense that for each vector $w$ of $\Sigma$ the norm of $w$ is
positive with respect to the intersection form (\ref{intf}).
Rotating $\Sigma$ inside ${H}^2(S,\mathbb{R})$ corresponds to changing the Ricci-flat metric on $S$, while rotating ($j,\omega$) inside $\Sigma$ corresponds to changing the complex structure of $S$ keeping the metric invariant. This is in accordance with Yau's theorem which guarantees
the existence of a unique Ricci-flat metric given a choice of complex structure
and the cohomology class of the K\"{a}hler form.
 The examples of K3$\times$K3 manifolds we will discuss later in this section are obtained by specifying within $H^2(S,\mathbb{R})$ a triplet of mutually-orthogonal, real, spacelike vectors ($j,\mathrm{Re}\omega,\mathrm{Im}\omega$) thus fixing the complex structure and the cohomology class of the K\"{a}hler form of $S$ (hence also fixing the Ricci-flat metric), and similarly for $\tilde{S}$.

To be more concrete, let us take the example of K3 surfaces with maximal Picard number, $\rho=20$,\footnote{K3 surfaces with
$\rho=20$ are sometimes called `singular K3 surfaces' in the mathematics literature \cite{si}, although they are in fact smooth manifolds.} where $\rho$ is defined as the rank of the Picard lattice $\mathrm{Pic}(S)=H^2(S,\mathbb{Z})\cap H^{1,1}(S)$. For $S$ an algebraic K3 surface, it can be seen that $\mathrm{Pic}(S)$ has signature $(1,\rho-1)$. Let us also define the transcendental lattice $T_S$ of $S$ as the orthogonal complement of $\mathrm{Pic}(S)$ in $H^2(S,\mathbb{Z})$; $T_S$ has signature $(2,\rho-20)$.
We will use the following, see e.g. \cite{morr}.

{\bf Lemma}: If $\Lambda\subset \Gamma$ is a primitive sublattice of $\Gamma$ (i.e. such that the quotient $\Gamma/\Lambda$ is a free abelian group) of signature $(1,\rho-1)$ then there exists an algebraic K3 surface $S$ such that $\mathrm{Pic}(S)=\Lambda$.

Consider the lattice $\Lambda$ generated by the basis vectors
$\{ e_{1,1}, e_{2,1}, e_{-,2}, e_{-,3}, e_i\}$, i=7,\dots 22, where we use the same enumeration as below (\ref{intf}). Clearly $\Lambda$ is a sublattice
of $\Gamma$. Its intersection form is given by
$$
\left( \begin{array}{cc}
-2&0\\
0&-2
\end{array} \right)\oplus U\oplus(-\mathfrak{e}_8)\oplus(-\mathfrak{e}_8)
$$
and so $\Lambda$ has signature (1,19). Moreover it can easily be seen that
 $\Gamma/\Lambda\cong\mathbb{Z}\oplus\mathbb{Z}$. Hence all the conditions of the lemma cited above are satisfied (for $\rho=20$) and there exists  an algebraic K3 surface $S$ with $\mathrm{Pic}(S)=\Lambda$. The transcendental lattice $T_S$ is generated by $\{e_{+2},e_{+3}\}$ and has signature (2,0).\footnote{Note that $\mathrm{Pic}(S)\oplus T_S$ is not in general equal to $\Gamma$. Indeed let $L$ be a unimodular lattice and $M$ a primitive sublattice of $L$. Then $M\oplus M^{\perp}=L$ if and only if $M$ is unimodular \cite{latt}. In the case at hand, $\Lambda$ is not unimodular.}
Since the K3 surface has maximal Picard number we have the isomorphisms $H^{2,0}(S)\oplus H^{0,2}(S)\cong T_S\otimes\mathbb{C}$ and $H^{1,1}(S,\mathbb{R})\cong \Lambda\otimes\mathbb{R}$. The holomorphic two-form $\omega$ is then proportional to $e_{+2}+ie_{+3}\in T_S\otimes\mathbb{C}$, while the (cohomology class of the) K\"{a}hler form $j$
defines a spacelike direction in $\Lambda\otimes\mathbb{R}$ and
 may be taken proportional to $e_{+1}$. All three of our explicit examples in this section use this simple choice of $j$, $\omega$. More precisely,
we take
\eq{\spl{
j=\sqrt{2\pi v}e_{+1}~&;~~~\omega=\sqrt{2\pi v}(e_{+2}+ie_{+3})\\
\tilde{j}=\sqrt{2\pi\tilde{v}}\tilde{e}_{+1}~&;~~~\tilde{\omega}=
\sqrt{2\pi\tilde{v}}(\tilde{e}_{+2}+i\tilde{e}_{+3})
~,}}
with $v$, $\tilde{v}\in\mathbb{R}_+$, so that (\ref{k3}) and its analogue for $\tilde{S}$ are satisfied with
\eq{\int_{S}\mathrm{vol_4}=2\pi v~;~~~
\int_{\tilde{S}}\mathrm{\widetilde{vol}_4}=2\pi\tilde{v}~.}

$\bullet$ Example 1 with $\mathcal{N}=1$

We set $C=2A$, $B=0$, $f^{\alpha\beta}=0$ in (\ref{k9}) and we take:
\eq{C = \pm \frac{2}{\sqrt{v \tilde{v}}}
~.}
To check the flux quantization condition (\ref{qk3}) we note first that
a basis of $H_4(S\times\tilde{S},\mathbb{Z})$ is given by $S$, $\tilde{S}$ and
all four-cycles of the form $\mathcal{C}_2\times\tilde{\mathcal{C}}_2$, where
$\mathcal{C}_2$ is a basis of $H_2(S,\mathbb{Z})$ and $\tilde{\mathcal{C}}_2$ is a basis of $H_2(\tilde{S},\mathbb{Z})$.
It can then be seen that (\ref{qk3}) is satisfied, with the only non-zero charges given by
\eq{
\frac{1}{2\pi}\int_{\mathcal{C}_{a,1}\times\mathcal{C}_{b,1}}G=\pm 2~;~~~
\frac{1}{2\pi}\int_{\mathcal{C}_{a,2}\times\mathcal{C}_{b,2}}G=\pm 1~;~~~
\frac{1}{2\pi}\int_{\mathcal{C}_{a,3}\times\mathcal{C}_{b,3}}G=\mp 1~,
}
for $a,b=1,2$. The topological constraint (\ref{tadpole2}) is satisfied for $N_{M2}=12$, where we have taken into account that $\chi(S\times\tilde{S})=24^2$.

$\bullet$ Example 2 with $\mathcal{N}=2$

We set $A=0$, $B=0$, $f^{\alpha\beta}=0$ in (\ref{k9}) and  we take
\eq{C = \pm \frac{2}{\sqrt{v \tilde{v}}}
~,}
so that the topological constraint (\ref{tadpole2}) is satisfied without M2 branes,  $N_{M2}=0$. Moreover it can be seen that (\ref{qk3}) is satisfied, with the only non-zero charges given by
\eq{
\frac{1}{2\pi}\int_{\mathcal{C}_{a,1}\times\mathcal{C}_{b,1}}G=\pm2~;~~~
\frac{1}{2\pi}\int_{S}G=\mp4\sqrt{\frac{v}{\tilde{v}}}
~;~~~\frac{1}{2\pi}\int_{\tilde{S}}G=\mp4\sqrt{\frac{\tilde{v}}{v}}
~,}
for $a,b=1,2$. A solution to the charge quantization constraints is thus obtained by setting
\eq{
v = 4^n \tilde{v}
~,}
for $n=0,\pm1,\pm2$.
This example possesses enhanced $\ncal=2$ supersymmetry since the four-flux is of the form (\ref{l2}).

$\bullet$ Example 3 with $\mathcal{N}=4$

We set $A=C$, $B=0$, $f^{\alpha\beta}=0$ in (\ref{k9}) and we take
\eq{C = \pm \frac{1}{\sqrt{v \tilde{v}}}
~.}
It can then be seen that (\ref{qk3}) is satisfied, with the only non-zero charges given by
\eq{\spl{
\frac{1}{2\pi}\int_{\mathcal{C}_{a,1}\times\mathcal{C}_{b,1}}G=\pm1~&;~~~
\frac{1}{2\pi}\int_{\mathcal{C}_{a,2}\times\mathcal{C}_{b,2}}G=\pm1~;~~~
\frac{1}{2\pi}\int_{\mathcal{C}_{a,3}\times\mathcal{C}_{b,3}}G=\mp1\\
&\frac{1}{2\pi}\int_{S}G=\frac{1}{2\pi}\int_{\tilde{S}}G=\pm2
~,}}
for $a,b=1,2$. The topological constraint (\ref{tadpole2}) is satisfied for $N_{M2}=14$. This example possesses enhanced $\ncal=4$ supersymmetry as it is a special case of the last example discussed in section \ref{sec:enhancement}.

\section{Conclusions}\label{secconcl}

We have presented  supersymmetric backgrounds with two real supercharges of
M-theory and type IIA supergravity (with non-zero Romans mass in general)
on eight-dimensional manifolds with $SU(4)$ structure, limiting our analysis to a strict  $SU(4)$ ansatz for the Killing spinors.
The present paper may be viewed as a continuation of \cite{pt} where
we considered $\ncal=(2,0)$ IIB supergravity on eight-dimensional manifolds with strict $SU(4)$ structure.
We hope to return to the generalization to dynamic $SU(4)$ ans\"{a}tze in the future. Since two-dimensional $\ncal=(2,0)$ supersymmetry is incompatible with a strict  $SU(4)$ ansatz on the IIA side, generalizing to dynamic $SU(4)$ ans\"{a}tze is also necessary in order to be able  to use the conjecture of \cite{patalong,pt}, recently proved in \cite{rosa}, on the correspondence between $\ncal=(2,0)$ supersymmetry of type II supergravity on eight-dimensional manifolds and generalized calibrations.

By taking the `internal' manifold to be an appropriate non-compact Calabi-Yau fourfold, our results could be used to construct novel explicit examples of backgrounds with two real supercharges which can be thought of as higher-dimensional analogues of the Klebanov-Strassler geometry, along the lines of \cite{clp,kp}. Such warped $\mathbb{R}^{1,d}\times\mcal_8$, $d=1,2$, backgrounds may admit an interpreation as gravity duals of certain $(d+1)$-dimensional field theories. We would like to return to this question in the future.

In the same way that the $\ncal=1$ Calabi-Yau flux vacua of section \ref{sec:cy} may be understood as spontaneous supersymmetry-breaking vacua of three-dimensional  $\ncal=2$ gauged supergravity \cite{haacksamt}, it would be interesting to have a description of the Calabi-Yau limit of the $\ncal=(1,1)$ solutions presented in section \ref{medium} as partial supersymmetry-breaking vacua of the two-dimensional $\ncal=(2,2)$ supergravity theories obtained by Kaluza-Klein reduction of IIA string theory on Calabi-Yau fourfolds \cite{gateswitten,hg}.

It appears that none of the explicit K3$\times$K3 flux vacua presented in section \ref{sec:examples} corresponds to an  F-theory vacuum with four-dimensional Poincar\'e invariance, as follows from the form of the four-flux in these examples. For the reasons already mentioned in the introduction, it would be interesting to search for explicit examples of the $\ncal=1$ M-theory vacua presented here that possess  F-theory duals. These duals would presumably be non-supersymmetric as four-dimensional vacua but should be related to supersymmetric three-dimensional vacua upon circle compactification, along the lines of \cite{witten}.

\section*{Acknowledgment}

We would like to thank Henning Samtleben for useful discussions.

\appendix

\section{Spinor and gamma matrix conventions}\label{app1}

For a spinor $\psi$ in any dimension we define:
\eq{\widetilde{\psi}\equiv\psi^{Tr}C^{-1}~,}
where $C$ is the charge conjugation matrix. In Lorentzian signatures, we also define
\eq{\overline{\psi}\equiv\psi^{\dagger}\G^{0}~.}
In all dimensions the Gamma matrices are taken to obey
\eq{
(\G^M)^{\dagger}=\G^0\G^M\G^0~,
}
where the Minkowski metric is mostly plus. Antisymmetric products of
Gamma matrices are defined by
\eq{
\G^{(n)}_{M_1\dots M_n}\equiv\G_{[M_1}\dots\G_{M_n]}~.
}

\subsection*{Two Lorentzian dimensions}

The charge conjugation matrix in $1+1$ dimensions satisfies
\eq{
C^{Tr}=-C; ~~~~~~ (C\g^\mu)^{Tr}=C\g^\mu; ~~~~~~ C^*=-C^{-1}~.
}
The fundamental (one-dimensional, chiral) spinor representation is real.
In this paper we work with a complexified chiral spinor $\zeta$ (i.e. one complex degree of freedom). We define:
\eq{
\zeta^c\equiv\g_0C\zeta^*~.
}
The chirality matrix is defined by
\eq{
\g_3\equiv{{-}}\g_0\g_1 ~.
}
The Hodge-dual of an antisymmetric product of
gamma matrices is given by
\eq{
\star\g_{(n)}\g_3={{- (-1)^{\frac{1}{2}n(n+1)}}}\g_{(2-n)}~.
\label{hodge2}
}

\subsection*{Eight Euclidean dimensions}

The charge conjugation matrix in $8$ dimensions satisfies
\eq{
C^{Tr}=C; ~~~~~~ (C\g^\mu)^{Tr}=C\g^\mu; ~~~~~~ C^*=C^{-1}~.
}
The fundamental (eight-dimensional, chiral) spinor representation is real.
In this paper we work with a complexified chiral spinor $\eta$ (i.e. eight complex degrees of freedom). We define:
\eq{
\eta^c\equiv C\eta^*~.
}
The chirality matrix is defined by
\eq{
\g_9\equiv\g_1\dots\g_8 ~.
}
The Hodge-dual of an antisymmetric product of
gamma matrices is given by
\eq{
\star\g_{(n)}\g_9=(-)^{\frac{1}{2}n(n+1)}\g_{(8-n)}~.
\label{hodge8}
}

\subsection*{Ten Lorentzian dimensions}

The charge conjugation matrix in $1+9$ dimensions satisfies
\eq{
C^{Tr}=-C; ~~~~~~ (C\G^M)^{Tr}=C\G^M; ~~~~~~ C^{*}=-C^{-1}~.
}
The fundamental (16-dimensional, chiral) spinor representation
$\e$ is real, where we define the reality condition by
\eq{
\overline{\e}=\widetilde{\e}~.
}
The chirality matrix is defined by
\eq{
\G_{11}\equiv{{-}}\G_0\dots\G_9 ~.
}
We decompose the ten-dimensional Gamma matrices as
\eq{
\left\{ \begin{array}{ll}
\G^{\mu}=\g^\mu\otimes \bbone &, ~~~~~\mu=0, 1\nn\\
\G^m=\g_3\otimes \g^{m-1} &, ~~~~~ m=2\dots 9
\end{array} \right.
~.}
It follows that
\eq{
C_{10}=C_2\otimes C_8;~~~~~~ \G_{11}=\g_3\otimes\g_9~.
}
The Hodge-dual of an antisymmetric product of
gamma matrices is given by
\eq{
\star \G_{(n)} \G_{11}= {{- (-1)^{\frac{1}{2}n(n+1)}}}      \G_{(10-n)}~.
}

\section{$SU(4)$ structures}\label{sec3}

As we will now review a nowhere-vanishing complex, chiral, pure spinor $\eta$ of unit norm in eight euclidean dimensions defines an $SU(4)$ structure. In eight euclidean dimensions not every complex chiral spinor is pure: the property of purity is equivalent to the condition
\eq{\label{p1}
\we\eta=0~.}
Let $\eta_R$, $\eta_I$ be the real, imaginary part of $\eta$ respectively.
We will impose the normalization:
\eq{\label{b2}
\eta=\frac{1}{\sqrt{2}}(\eta_R+i\eta_I)~;~~~~~\we_R\eta_R=\we_I\eta_I=1~,}
so that $\tilde{\eta}^c\eta=1$, and (\ref{p1}) is equivalent to $\eta_R$, $\eta_I$ being orthogonal to each other: $\we_R\eta_I=\we_I\eta_R=0$.

Let us  define a real two-form $J$ and a complex self-dual four-form $\Omega$ through the spinor bilinears
\eq{\spl{\label{b3}
iJ_{mn}&=\widetilde{\eta^c}\g_{mn}\eta\\
\Omega_{mnpq}&=\we\g_{mnpq}\eta~.
}}
It can then be shown by Fierzing that these forms obey:
\eq{\spl{\label{b4}
J\wedge\Omega&=0\\
\frac{1}{16}\Omega\wedge\Omega^*&=\frac{1}{4!}J^4 =\mathrm{vol}_8~,
}}
up to a choice of orientation,
and hence define an $SU(4)$ structure.
The reduction of the structure group  can alternatively be seen from the fact that $Spin(6)\cong SU(4)$ is the stabilizer inside $Spin(8)$ of the pair of orthogonal Majorana-Weyl unit spinors $\eta_R$, $\eta_I$.

Raising one index of $J$ with the metric defines an almost complex structure:
\eq{
J_m{}^pJ_p{}^n=-\delta_m^n
~.}
Using the almost complex structure
we can define the projectors
\eq{
(\Pi^{\pm})_m{}^n\equiv\frac{1}{2}(\delta_{m}{}^{n}\mp i J_m{}^n)
~,}
with respect to which $\Omega$ is holomorphic
\eq{
(\Pi^{+})_m{}^i\Omega_{inpq}=\Omega_{mnpq}~; ~~~~~(\Pi^{-})_m{}^i\Omega_{inpq}=0 ~.
}
Further useful relations are given in appendix \ref{app2}.

In eight dimensions the Clifford algebra $Cl(8)$ is equivalent to the set $\mathbb{R}[16]$ of real $16\times 16$ matrices. With complex coefficients the
gamma matrices generate $Cl(8)\otimes\mathbb{C}\cong\mathbb{C}[16]$. Since a complex Dirac spinor in eight dimensions can be thought of as a vector of $\mathbb{C}^{16}$ and $\mathbb{C}[16]$ acts transitively on $\mathbb{C}^{16}-\{0\}$, any complex Dirac spinor can be expressed as an element of $Cl(8)$, with complex coefficients, acting on the non-vanishing spinor $\eta$.

More explicitly, let $\xi_+$, $\xi_-$ be arbitrary positive-, negative-chirality complexified spinors  respectively. Using equations (\ref{fierzsu}), it follows from the previous paragraph that $\xi_\pm$ can be expressed as:
\eq{\spl{
\xi_+&=\varphi~\!\eta+\chi~\!\eta^c+\varphi_{mn}\g^{mn}\eta^c\\
\xi_-&={{\lambda_{m}}} \g^{m}\eta+\chi_{m}\g^{m}\eta^c
~,
\label{xiexp}}}
where $\varphi$, $\chi$ are complex scalars, $\varphi_m$, $\chi_m$ are complex
(0,1)-, (1,0)-forms respectively and $\varphi_{mn}$ is a complex (2,0)-form. As a consistency check, we note that the arbitrary positive-chirality spinor $\xi_+$ is parametrized by eight complex degrees of freedom: two complex d.o.f.s from the complex scalars $\varphi$, $\chi$ plus six complex d.o.f.s from the complex (2,0)-form  $\varphi_{mn}$. Similarly $\xi_-$ is parametrized by four plus four complex d.o.f.s coming from the (0,1)-, (1,0)-forms $\varphi_m$, $\chi_m$ respectively.

\subsection{Torsion classes}\label{sectorsion}

The intrinsic torsion $\tau$ (see e.g. \cite{gauntlett} for a review) transforms in the $\Lambda^1(\mcal_8)\otimes su(4)^\perp$, where $su(4)^\perp$ is the complement of the adjoint of $su(4)$ inside the adjoint of $so(8)$. It follows that
\eq{\spl{
\tau&\in(\bf{4}\oplus\bf{\bar{4}})\otimes(\bf{1}\oplus \bf{6}\oplus\bf{6})\nn\\
&\sim(\bf{4}\oplus\bf{\bar{4}})\oplus(\bf{20}\oplus\bf{\bar{20}})
\oplus(\bf{20}\oplus\bf{\bar{20}})
\oplus(\bf{4}\oplus\bf{\bar{4}})\oplus(\bf{4}\oplus\bf{\bar{4}})
~,}}
in a decomposition in terms of irreps of $su(4)$. We then decompose $\tau$ in five torsion classes $W_1,\dots, W_5$, according to the second line on the right-hand side above. These torsion classes are the obstructions to the closure of the forms $J$, $\Omega$. Explicitly we will choose the following
parameterization:\footnote{\label{f16}We define the contraction
between a $p$-form $\varphi$ and a $q$-form $\chi$, $p\leq q$, by
\eq{
\varphi\lrcorner\chi=\frac{1}{p!(q-p)!}\varphi^{m_1\dots m_p}
\chi_{m_1\dots m_p n_1\dots n_{q-p}}\d x^{n_1}\wedge\dots\wedge\d x^{n_{q-p}}
~.\nn}
Once the normalization of the $W_1$ term on the right-hand side of the first equation in (\ref{torj}) is fixed, the $W_1$ term on the right-hand side of the second equation can be determined as follows: Starting from $\d(J\wedge\Omega)=0$ we substitute for $\d J$, $\d\Omega$ using (\ref{torj}), taking (\ref{jvol}), (\ref{omvol}) into account and noting that
$W_2\wedge J\wedge J=0$ since $W_2$ is primitive.}
\eq{\spl{\label{torj}
\d J&=W_1\lrcorner\Omega^*+W_3+W_4\wedge J+\mathrm{c.c.}\\
\d\Omega&=\frac{8i}{3}W_1\wedge J\wedge J+W_2\wedge J+W^*_{5}\wedge\Omega
~,}}
where $W_1$,  $W_4$, $W_5\sim\bf{4}$ are complex (1,0)-forms and $W_2$, $W_3\sim\bf{20}$ are complex traceless (2,1)-forms.

Equivalently, the torsion classes are the obstructions to the spinor $\eta$ being covariantly constant with respect to the Levi-Civita connection. Explicitly we have:
\eq{\spl{\label{toreta}
\nabla_m \eta
&= \left( \frac{3}{4}W_{4m} - \frac{1}{2} W_{5m} - \mathrm{c.c.}\right) \eta +\frac{i}{24} \O^*_{mnkl}  W_{1}^{n} \g^{kl}\eta\\
& +\left(
-\frac{i}{16} W_{2mkl}
- \frac{1}{32}  \O_{mnkl} W_{4}^{n*}
+ \frac{i}{64} W^*_{3mnp} \O^{np}{}_{kl}
 \right) \g^{kl}\eta^c~.
}}
This can be seen as follows. From the discussion around (\ref{xiexp}) and the fact that $\nabla_m\eta$ transforms in the $\bf{8}\otimes\bf{8^+}$ of $so(8)$, we can expand
\eq{\label{toretaprov}
\nabla_m\eta=\varphi_m \eta + {{\vartheta_m}} \eta^c + \Psi_{m,pq} \Omega^{pqrs} \gamma_{rs} \eta^c~,}
for some complex coefficients $\varphi_m$, $\vartheta_m\sim{\bf 4}\oplus{\bf \bar{4}}$, $\Psi_{m,pq}\sim({\bf 4}\oplus{\bf \bar{4}})\otimes{\bf 6}$. Furthermore we decompose:
\begin{align}
\Psi_{m,pq} &= \Omega^*_{mpqr}A^r +  \Omega_{pq}^{*\phantom{pq}rs} \tilde{\varphi}_{rsm}+ (\Pi^+)_{m[p}B^*_{q]} + (\Pi^+)_{m}^{\phantom{m}n} \psi^*_{npq}
\end{align}
where $A,B \sim{\bf 4}$ are complex (1,0)-forms and $\tilde{\varphi},\psi \sim{\bf 20}$ are complex traceless (2,1)-forms.
Multiplying (\ref{toretaprov}) on the left with $\widetilde{\eta^c}\g_{ij}$ and $\we\g_{ijk}$, antisymmetrizing in all indices  in order to form $\d J$ and $\d \Omega$ respectively as spinor bilinears and  comparing with (\ref{torj}) then leads to \eqref{toreta}.

As can be seen from \eqref{torj}, the obstruction to having an integrable  almost complex structure is given by $W_1$, $W_2$. Conversely, if  $W_1$, $W_2$ vanish one can use \eqref{toreta} to show that the Nijenhuis tensor vanishes and thus the almost complex structure is integrable.

\subsection{Tensor decomposition}\label{sec4}

Under an $so(8)\rightarrow su(4)$ decomposition the two-form, the three-form, the self-dual and the anti-self-dual four-form of $so(8)$ decompose respectively as:
\eq{\spl{
\bf{ 28}&\rightarrow \bf{ (6\oplus {6})\oplus(1\oplus 15) }\nn\\
\bf{ 56}&\rightarrow \bf{ (4\oplus \bar{{4}})\oplus(4\oplus 20)\oplus(\bar{4}\oplus \bar{20}) }\nn\\
\bf{35^+}&\rightarrow \bf{(1\oplus 1)\oplus (6\oplus {6})\oplus 20'\oplus  1}\nn\\
\bf{35^-}&\rightarrow \bf{(10\oplus \bar{10})\oplus 15}~.
}}
Explicitly we decompose the RR tensors as follows.

{$\bullet$  Real two-form}
\eq{\label{dec2}F_{mn}=f^{(1,1)}_{2|mn}+f_2J_{mn}+\left(f^{(2,0)}_{2|mn}+\mathrm{c.c.}\right)~,}
where $f^{(1,1)}_{2|mn}\sim \bf{15}$ is a real traceless (1,1)-form, $f_2\sim \bf{1}$ is a real scalar, $f^{(2,0)}_{2|mn}\sim \bf{6}\oplus\bf{6}$ is a complex (2,0)-form. Note that given a complex (2,0)-form $\varphi^{(2,0)}$ transforming
in the reducible module $\bf{6}\oplus\bf{6}$, one may form irreducible representations thereof by imposing
a pseudoreality condition:
\eq{\label{pseudo}
\varphi^{(2,0)}_{mn}=\frac18~\! e^{i\theta}\O_{mn}{}^{pq}\varphi^{(0,2)}_{mn}
~,}
where $\theta\in S^1$ is an arbitrary phase.

{$\bullet$  Real three-form}
\eq{\label{dec3}F_{mnp}=f^{(2,1)}_{3|mnp}+3f^{(1,0)}_{3|[m}J_{np]}
+\tilde{f}^{(1,0)}_{3|s}\Omega^{s*}{}_{mnp}
+\mathrm{c.c.}~,}
where $f^{(2,1)}_{3|mnp}\sim \bf{20}$ is a complex traceless (2,1)-form, ${f}^{(1,0)}_{3|m}, \tilde{f}^{(1,0)}_{3|m}\sim \bf{4}$ are complex (1,0)-forms.

{$\bullet$  Real self-dual four-form}
\eq{\label{dec4s}F^+_{mnpq}=
f^{(2,2)}_{4|mnpq}+6f_{4}J_{[mn}J_{pq]}
+\left(6f^{(2,0)}_{4|[mn}J_{pq]}+\tilde{f}_4\Omega_{mnps}
+\mathrm{c.c.}\right)~,}
where $f^{(2,2)}_{4|mnpq}\sim \bf{20'}$ is a real traceless (2,2)-form,  $f_4\sim \bf{1}$ is a real scalar,
$f^{(2,0)}_{4|mn}\sim \bf{6}+\bf{6}$ is a complex (2,0)-form,
$\tilde{f}_4\sim (\bf{1}\oplus\bf{1})$ is a complex scalar.

{$\bullet$  Real anti self-dual four-form}
\eq{\label{dec4a}F^-_{mnpq}=6f^{(1,1)}_{4|[mn}J_{pq]}
+\left(f^{(3,1)}_{4|mnpq}+\mathrm{c.c.}\right)~,}
where $f^{(1,1)}_{4|mn}\sim \bf{15}$ is a real traceless (1,1)-form,
$f^{(3,1)}_{4|mnpq}\sim \bf{\bar{10}}$ is a complex traceless (3,1)-form.
The following identity can easily be shown:
\eq{
\Omega^*_{[s}{}^{mnp}f^{(3,1)}_{4|q]mnp}=0
~.}
For the RR-forms $F_p$ with $p>4$ we expand the Hodge duals $\star_8F_p$  as above. For the real six-form:
\eq{\label{dec6}(\star_8F_6)_{mn}=
f^{(1,1)}_{6|mn}+f_6J_{mn}+\left(f^{(2,0)}_{6|mn}+\mathrm{c.c.}\right)~,}
and similarly for the eight-form,
\eq{\label{dec8}\star_8F_8=
f_{8}~.}
For the NSNS three-form $H$ we decompose similarly:
\eq{\label{b19}
H=e^{2A}\mathrm{vol}_2\wedge h_1+h_3
~,}
where as before $\mathrm{vol}_2$ is the unwarped volume
element of $\mathbb{R}^{1,1}$; $h_1$, $h_3$ are real one-, three-forms
on $\mathcal{M}_8$ respectively. These further decompose to irreducible
$su(4)$-modules:
\eq{h_{1|m}=h_{1|m}^{(1,0)}+\mathrm{c.c.}~,}
with ${h}^{(1,0)}_{1|m}\sim \bf{4}$ a complex (1,0)-form, and
\eq{
h_{3|mnp}=
h^{(2,1)}_{3|mnp}+3h^{(1,0)}_{3|[m}J_{np]}
+\tilde{h}^{(1,0)}_{3|s}\Omega^{s*}{}_{mnp}
+\mathrm{c.c.}~,}
where $h^{(2,1)}_{3|mnp}\sim \bf{20}$ is a complex traceless (2,1)-form, ${h}^{(1,0)}_{3|m}, \tilde{h}^{(1,0)}_{3|m}\sim \bf{4}$ are complex (1,0)-forms.

\subsection{Useful formul\ae{}}\label{app2}

The following useful identities can be proved  by Fierzing \cite{tsim}:
\eq{\spl{
\frac{1}{4!\times 2^4}~&\Omega_{rstu}\Omega^{*rstu}=1\\
\frac{1}{6\times 2^4}~&\Omega_{irst}\Omega^{*mrst}
=(\Pi^+)_{i}{}^{m}\\
\frac{1}{4\times 2^4}~&\Omega_{ijrs}\Omega^{*mnrs}
=(\Pi^+)_{[i}{}^{m}(\Pi^+)_{j]}{}^{n}\\
\frac{1}{6\times 2^4}~&\Omega_{ijkr}\Omega^{*mnpr}
=(\Pi^+)_{[i}{}^{m}(\Pi^+)_{j}{}^{n}(\Pi^+)_{k]}{}^{p}\\
\frac{1}{4!\times 2^4}~&\Omega_{ijkl}\Omega^{*mnpq}
=(\Pi^+)_{[i}{}^{m}(\Pi^+)_{j}{}^{n}(\Pi^+)_{k}{}^{p}(\Pi^+)_{l]}{}^{q}~,
\label{bfive}
}}
Moreover, we have
\eq{\spl{
\widetilde{\eta^c}\eta=1; &~~~~~\we\eta=0\\
\widetilde{\eta^c}\g_{mn}\eta=iJ_{mn}; &~~~~~\we\g_{mn}\eta=0\\
 \widetilde{\eta^c}\g_{mnpq}\eta=-3J_{[mn}J_{pq]}; &~~~~~\we\g_{mnpq}\eta=\Omega_{mnpq}\\
\widetilde{\eta^c}\g_{mnpqrs}\eta=-15iJ_{[mn}J_{pq}J_{rs]}; &~~~~~\we\g_{mnpqrs}\eta=0\\
\widetilde{\eta^c}\g_{mnpqrstu}\eta=105J_{[mn}J_{pq}J_{rs}J_{tu]}; &~~~~~\we\g_{mnpqrstu}\eta=0 ~,
\label{usefids}
}}
where we have made use of the identities
\eq{\spl{\label{jvol}
\sqrt{g}\; \varepsilon_{mnpqrstu}J^{rs}J^{tu}&=24J_{[mn}J_{pq]}\\
\sqrt{g} \;\varepsilon_{mnpqrstu}J^{tu}&=30J_{[mn}J_{pq}J_{rs]}\\
\sqrt{g} \;\varepsilon_{mnpqrstu}&=105J_{[mn}J_{pq}J_{rs}J_{tu]}
~.
}}
Note that the bilinears
$\we\g_{(p)}\eta$, ~$\widetilde{\eta^c}\g_{(p)}\eta$, vanish for $p$ odd.
The last line of equation (\ref{bfive}) together with the last line of the
equation above imply
\eq{\label{omvol}
\Omega_{[ijkl}\Omega^*_{mnpq]}=\frac{8}{35}\sqrt{g}\; \varepsilon_{ijklmnpq}~.
}
%
%
%
Finally, the following relations are useful in the analysis of the Killing spinor equations:
\eq{\spl{
\g_m\eta&=(\Pi^+)_{m}{}^{n}\g_n\eta\\
\g_{mn}\eta&=iJ_{mn}\eta -\frac{1}{8}\Omega_{mnpq}\g^{pq}\eta^c   \\
\g_{mnp}\eta&=3iJ_{[mn}\g_{p]}\eta
-\frac{1}{2}\Omega_{mnpq}\g^q\eta^c\\
\g_{mnpq}\eta&=-3J_{[mn}J_{pq]}\eta -\frac{3i}{4}J_{[mn}\Omega_{pq]ij}\g^{ij}\eta^c
+\Omega_{mnpq}\eta^c
~.
\label{fierzsu}
}}
The action of $\gamma_{m_1\dots m_p}$, $p\geq 5$, on $\eta$ can be related to
the above formul{\ae}, using
the Hodge properties of gamma matrices given in appendix \ref{app1}.

\subsection*{Formul\ae{} needed for the dilatino equations}

In order to solve the dilatino equations, we make use of the following:
\eq{\spl{\label{c4}
{{F}}_0 \eta &= f_0 \eta \\
\underline{F_2} \eta &= 4 i f_2 \eta - \frac{1}{16} f^{(0,2)}_{2|mn} \Omega^{mnpq}\gamma_{pq} \eta^c \\
\underline{F_4} \eta &= -12 f_4 \eta + 16\tilde{f}^{*}_{4} \eta^c - \frac{i}{8} f{^{(0,2)}_{4|mn}} \Omega^{mnpq} \gamma_{pq} \eta^c~,
}}
and similarly for the $p$-forms $F_p$ with $p>4$. Moreover
\eq{\spl{
\underline{h_1} \eta &= h_{1|m}^{(0,1)} \g^m \eta \\
\underline{h_3} \eta &= 3i h_{3|m}^{(0,1)} \g^m \eta
+8 \tilde{h}_{3|m}^{(1,0)} \g_m \eta^c
~.}}

\subsection*{Formul\ae{} needed for the internal gravitino equations}
For the gravitino equations with $M=m$, we require
\eq{
\underline{h_{3|m}}  \eta = 3i ( h^{(1,0)}_{3|m} + h^{(0,1)}_{3|m}) \eta -
\left(\frac{i}{8}h^{(0,1)}_{3|n}\O^{\phantom{m}nrs}_m + \frac{1}{16} h^{(1,2)}_{3|mpq} \O^{pqrs}\right)\g_{rs} \eta^c
-\frac{1}{2}\tilde{h}^{(1,0)}_{3|n}\O^{*\phantom{m}npq}_m\g_{pq} \eta
}
and

\eq{\spl{\label{c7}
  {F}_0 \g_m \eta &= f_0   \g_m \eta\\
  \underline{F}_2 \g_m \eta &=  \left(2i f_2 \g_m - 2 f_{2|mn}^{(1,1)}\g^n \right) \eta - \frac{1}{4}   f_{2|np}^{(0,2)} \g_q \O_m^{\phantom{m}npq} \eta^c\\
  \underline{F}_4 \g_m \eta &=
- 4i f_{4|mn}^{(1,1)}\g^n  \eta
+ \frac{1}{6}  f_{4|mnpq}^{(1,3)} \g_r \O^{npqr} \eta^c~,
}}
and similarly for the $p$-forms $F_p$ with $p>4$.

\section{$\ncal=(2,2)$}\label{maximal}

We now consider the case where we have $(2,2)$ supersymmetry. This means that we consider two Majorana-Weyl spinors $\zeta_{1+}, \zeta_{2+}$ and take as our Killing spinor ansatz\footnote{This parameterization of the Killing spinor ansatz is slightly redundant. We have used it for the purpose of facilitating the comparison with the $\ncal=(1,1)$ case.}
\eq{\spl{\label{c1}
\epsilon_1 &=\frac{\alpha}{\sqrt{2}}\zeta_{1+}\otimes (\eta + \eta^c)
 + \frac{\b}{\sqrt{2}}\zeta_{2+}\otimes\left( e^{i \t_2}  \eta + e^{- i \t_2} \eta^c\right)\\
\epsilon_2 &=  \frac{\g}{\sqrt{2}}\zeta_{1-}\otimes\left( e^{i \t}  \eta + e^{-i \t}  \eta^c \right)
            + \frac{\delta}{\sqrt{2}}\zeta_{2-}\otimes \left( e^{i \t_3}  \eta + e^{- i \t_3} \eta^c \right) \;,
}}
with $a, \b, \g, \delta, \t, \t_2, \t_3$ real functions on $\mathcal{M}_8$.
Inserting (\ref{c1}) into the supersymmetry equations leads to
\eq{\spl{
W&=0\\
 2\d A+h_{1}&=0\\
\left(2\underline{\d\phi}+\underline{h_1}+\underline{h_3}\right)( \eta +   \eta^c )&=0\\
\left(2\underline{\d\phi}+\underline{h_1}+\underline{h_3}\right) (e^{i \t_2} \eta + e^{- i \t_2} \eta^c)  &=0\\
\left(2\underline{\d\phi}+\underline{h_1}-\underline{h_3}\right) (e^{i \t} \eta + e^{- i \t} \eta^c )&=0\\
\left(2\underline{\d\phi}+\underline{h_1}-\underline{h_3}\right) (e^{i \t_3} \eta + e^{- i \t_3} \eta^c )&=0\\
\left(\nabla_m+\partial_m\log\alpha+\frac{1}{4}h_{1|m}
+\frac{1}{4}\underline{h_{3|m}}\right)                 ( \eta +  \eta^c ) &=0\\
\left(\nabla_m+\partial_m\log\b+\frac{1}{4}h_{1|m}
+\frac{1}{4}\underline{h_{3|m}}\right)                 ( e^{i \t_2} \eta + e^{- i \t_2} \eta^c ) &=0\\
\left(\nabla_m+\partial_m\log\g+\frac{1}{4}h_{1|m}
-\frac{1}{4}\underline{h_{3|m}}\right)                  (e^{i \t} \eta + e^{- i \t} \eta^c )&=0\\
\left(\nabla_m+\partial_m\log\delta +\frac{1}{4}h_{1|m}
-\frac{1}{4}\underline{h_{3|m}}\right)                  (e^{i \t_3} \eta + e^{- i \t_3} \eta^c)&=0
~,}}
and
\eq{\spl{
{\underline{\sigma(F)}}( \eta + \eta^c) &=0\\
{\underline{\sigma(F)}}( e^{i \t_2} \eta + e^{- i \t_2} \eta^c) &=0\\
\underline{\sigma(F)}\gamma_m ( \eta + \eta^c )            &=0\\
\underline{\sigma(F)}\gamma_m ( e^{i \t_2} \eta + e^{- i \t_2} \eta^c ) &=0\\
{\underline{F}}( e^{i \t} \eta + e^{- i \t} \eta^c ) &=0\\
{\underline{F}}(e^{i \t_3} \eta + e^{- i \t_3}  \eta^c ) &=0\\
\underline{F}\gamma_m ( e^{i \t} \eta +  e^{- i \t}\eta^c ) &=0\\
\underline{F}\gamma_m ( e^{i \t_3}\eta +  e^{-i \t_3}\eta^c ) &=0
~.}}
Setting $e^\t = \pm e^{\t_3}$, $e^{2i \t_2} = 1$ one recovers the $\mathcal{N}=(1,1)$ ansatz of section \ref{medium}. For $\t \neq \t_3$ or $e^{ 2i \t_2 } \neq 1$ we find that
the RR sector is given by
\eq{\spl{
f_8 = f_0 &= 6 f_4 \\
f_6 = f_2 &= 0 \\
\tilde{f}_4 &= 0\\
f_{2|mn}^{(2,0)} =  f_{6|mn}^{(2,0)} =  f_{4|mn}^{(2,0)} &= 0\\
f_{4|mn}^{(1,1)} &= 0\\
f_{6|mn}^{(1,1)} &= -f_{2|mn}^{(1,1)} \\
f_{4|mn}^{(3,1)} &= 0
~,}}
leaving $f_0$ and $f_{2|mn}^{(1,1)}$ as free parameters. The NSNS sector is given by
\eq{\spl{
W &= 0\\
W_1 = W_2 = W_3 = W_4 = W_5 &=0\\
\a &= e^{\frac{1}{2} A }\\
\b &= C_1 e^{\frac{1}{2} A }\\
\g &= C_2 e^{\frac{1}{2} A }\\
\delta &= C_3 e^{\frac{1}{2} A }\\
h_3 &= 0\\
 d \phi &=  d A = - \frac{1}{2} h_1\\
d\t = d\t_2 = d\t_3 &= 0
~,}}
where $C_i$, i=1,2,3, are real constants.

\section{$\ncal=(1,0)$}\label{minimal}

The $\mathcal{N}=(1,0)$ case is solved analogously to the $\mathcal{N}=(1,1)$ case. Our Killing spinor ansatz in this case
is given by
\eq{
\epsilon_1=\frac{\alpha}{\sqrt{2}}\zeta_+\otimes(\eta+\eta^c)~,~~~
\epsilon_2=0~.
}
The resulting equations are:
\eq{\spl{\label{d2}
W         &=0\\
 2\d A+h_{1  }&=0\\
\left(2\underline{\d\phi}+\underline{h_1}+\underline{h_3} \right)(\eta+\eta^c)&=0\\
\left(\nabla_m+\partial_m\log\alpha+\frac{1}{4}h_{1|m}
+\frac{1}{4}\underline{h_{3|m}}\right)(\eta+\eta^c)       &=0
~,}}
and
\eq{\spl{\label{d3}
{\underline{\sigma(F)}}\left(\eta+\eta^c \right) &=0\\
\underline{\sigma(F)}\gamma_m\left(\eta+\eta^c \right)&=0\\
~.}}
Again, the RR and NSNS equations decouple and $\mathcal{M}_2$ reduces to Minkowski space.
The solutions of the system (\ref{d2}), (\ref{d3}) are given by
\eq{\spl{
W &= 0\\
\alpha &= e^{ \frac12 A}\\
h_1 &= - 2\d A\\
\tilde{h}_{3|m}^{(1,0)} &=  \frac{1}{4} \p_m^+\left( A - \phi \right) - \frac{3}{8}  i h_{3|m}^{(1,0)}\\
W_1 &=  -\frac{3}{8} h_{3|m}^{(1,0)}  - \frac{1}{2}i W_5 - \frac{3}{4}i \p_m^+\left( A - \phi \right) \\
W_2 &= 2 W_3 - 2i h^{(2,1)}_3\\
W_4 &= \frac{2}{3} W_5 - i h_{3}^{(1,0)}
~,}}
with $A$, $\phi$, $h_{3}^{(1,0)}$, $h_3^{(2,1)}$, $W_5$ unconstrained,
and
\eq{\spl{
f_8 &= f_0\\
f_6 &= - f_2\\
\tilde{f}_4 &= \frac{3}{4} f_0 - \frac{1}{8}f_4 +  \frac{i}{2}f_2\\
f_{2|mn}^{(2,0)} &= -  f_{6|mn}^{(2,0)} \\
\left( f_{2|mn}^{(2,0)} - \frac{1}{8} \O_{mn}^{\phantom{mn}pq}f_{2|pq}^{(0,2)}\right)
&= -i \left( f_{4|mn}^{(2,0)} + \frac{1}{8} \O_{mn}^{\phantom{mn}pq}f_{4|pq}^{(0,2)}\right)\\
f_{4|mn}^{(1,1)} &= 0\\
f_{6|mn}^{(1,1)} &= -f_{2|mn}^{(1,1)} \\
f_{4|mn}^{(3,1)} &= 0
~,}}
with $f_0$, $f_2$, $f_4$, $f_{2|mn}^{(1,1)}$ unconstrained.

\section{\label{app:uplift}Eleven-dimensional uplift of IIA}

We here briefly review the eleven-dimensional uplift of (massless) IIA supergravity. For the purposes of this appendix we switch to the standard (i.e. non-democratic) formulation of IIA.
The starting point is the eleven-dimensional action,
\eq{
S  = \int \d^{11}x\sqrt{-g_{11}}\big( R_{11} - \frac{1}{48} G^2\big) - \frac{1}{6} \int A_3 \wedge G  \wedge G ~,
}
where $R_{11}$ is the scalar curvature of the eleven-dimensional metric
$g_{11}$ and
\eq{
G=\d {A}_3~.
}
Assuming the geometry has a $U(1)$ isometry, we can
put the metric in the form:
\eq{
\d s_{11}^2=e^{-{1\over 6}\phi}\d s_{10}^2+
e^{{4\over 3}\phi}(C_1+\d z)^2  ~,
}
where  $\phi$, $C_1$, $A_3$ and the metric in $\d s_{10}^2$ are
all assumed independent of the coordinate $z$.
We also reduce the 3-form potential,
\eq{
A_3=C_3+B \wedge \d z
~.}
The eleven-dimensional Lagrangian then reduces to the type IIA
theory in the Einstein frame:
\eq{
\mathcal{L}= R_{10}  -{1\over 2} (\partial \phi)^2
-{1 \over 4} e^{{3\over 2}\phi} F_2^2
-{1 \over 12}e^{-\phi} H^2
-{1 \over 48}e^{{1\over 2}\phi} F_4^2 +\mathrm{C.S.}
~,}
where $R_{10}$ is the scalar curvature of the ten-dimensional metric
in $\d s^2_{10}$ and
\eq{
F_4=\d C_3 - H \wedge C_1; \,\,\,\,\, H=\d B; \,\,\,\,\, F_2=\d C_1
~,}
so that
\eq{
G= F_4 + H \wedge (\d z+C_1)~.
}
Let us also note that the ten-dimensional Einstein metric is related to the string metric through
\eq{
\d s^2_{E}=e^{-{1\over 2}{\phi}}\d s^2_{\mathrm{str}} ~.
}

\end{document}